\pdfoutput=1

\documentclass[11pt]{article}

\usepackage[preprint]{acl}
\usepackage{enumitem}
\usepackage{makecell} 
\setcellgapes{1pt} 
\usepackage{booktabs}
\usepackage{tcolorbox}
\usepackage{listings}
\usepackage{adjustbox}
\usepackage{array}
\usepackage{amsmath}
\usepackage{multirow}
\usepackage{lmodern}
\usepackage{caption} 
\usepackage{lipsum}
\usepackage{stfloats} 
\usepackage{colortbl}
\usepackage{tabularx}
\usepackage{booktabs}
\usepackage[normalem]{ulem}
\usepackage{booktabs}
\usepackage{multirow}
\usepackage{adjustbox} 
\usepackage{times}
\usepackage{latexsym}

\usepackage[T1]{fontenc}
\usepackage[utf8]{inputenc}
\usepackage{microtype}
\usepackage{inconsolata}
\usepackage{graphicx}
\usepackage{subcaption}
\usepackage[font=small,labelfont=bf]{caption} 
\newcommand{\name}{AutoDCWorkflow}
\usepackage{pifont}
\usepackage{booktabs}
\usepackage{amssymb}
\usepackage{array}
\usepackage{xcolor}
\newcommand{\cmark}{\textcolor{green!60!black}{\ding{51}}}  
\newcommand{\revise}[1]{\textcolor{black}{#1}}

\title{\name: LLM-based Data Cleaning Workflow Auto-Generation and Benchmark}

\author{
  Lan Li\thanks{Equal contribution} \quad
  Liri Fang\footnotemark[1] \quad
  Bertram Lud\"{a}scher \quad
  Vetle I. Torvik \\
  University of Illinois Urbana-Champaign \\
  \texttt{\{lanl2, lirif2, ludaesch, vtorvik\}@illinois.edu}
}

\begin{document}


\newpage

\maketitle

\begin{abstract} \label{sec:abstract}
Data cleaning is a time-consuming and error-prone manual process even with modern workflow tools like OpenRefine. Here, we present AutoDCWorkflow, an LLM-based pipeline for automatically generating data-cleaning workflows. The pipeline takes a raw table coupled with a data analysis purpose, and generates a sequence of OpenRefine operations designed to produce a minimal, clean table sufficient to address the purpose. Six operations correspond to common data quality issues including format inconsistencies, type errors, and duplicates.

To evaluate \name, we create a benchmark with metrics assessing answers, data, and workflow quality for 142 purposes using 96 tables across six topics. The evaluation covers three key dimensions: (1) \textbf{Purpose Answer}: can the cleaned table produce a correct answer? (2) \textbf{Column (Value)}: how closely does it match the ground truth table? (3) \textbf{Workflow (Operations)}: to what extent does the generated workflow resemble the human-curated ground truth? Experiments show that Llama 3.1, Mistral, and Gemma 2 significantly enhance data quality, outperforming the baseline across all metrics. Gemma 2-27B consistently generates high-quality tables and answers, \revise{while Gemma 2-9B excels in producing workflows that resemble human-annotated versions.}

\end{abstract}

\section{Introduction} \label{sec:introduction}

\begin{figure}[!th] 
\centering 
\includegraphics[width=\columnwidth]{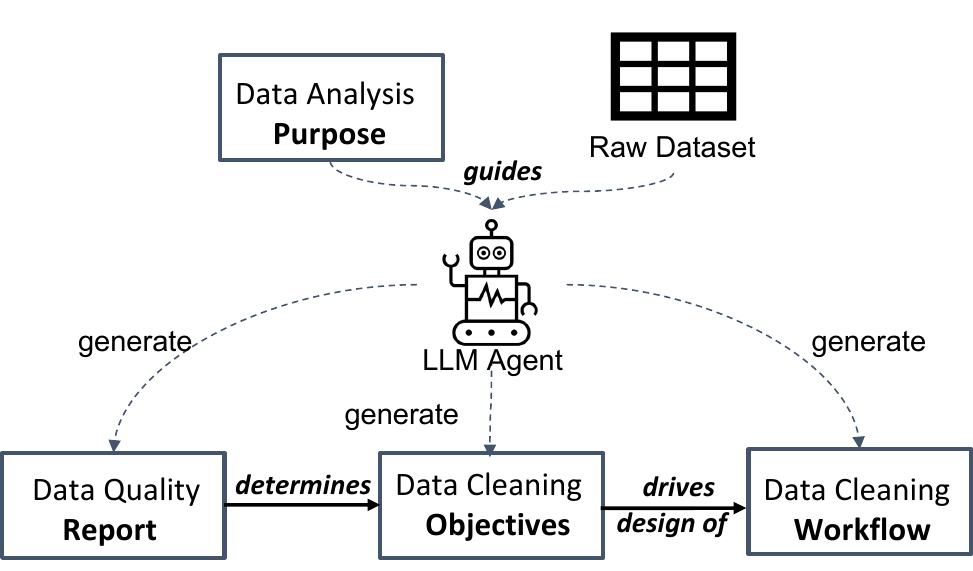}
\caption{In a purpose-driven data cleaning model, \name~ leverages LLM agents to decide each reasoning step and automatically generate a data cleaning workflow based on a curator-defined analysis purpose and a provided dirty dataset.
}
\vspace{-5pt}
\label{fig-llma-model}
\end{figure}

Data curation and cleaning are critical processes that prepare raw data for analysis, ensuring high quality and reliability ~\cite{chen2023seed}. Implementing a sequence of data operations in the form of a data cleaning workflow improves data quality, enabling downstream analyses and yielding trustworthy results and actionable insights~\cite{li2021automatic, wilkinson_fair_2016, parulian2022dcm, mcphillips2019reproducibility}. \revise{To facilitate automation and reuse, tools like OpenRefine~\cite{OpenRefine} and Trifacta \cite{kandel2011wrangler} automatically capture reusable, executable workflows that ensure consistent results when re-applied to the same raw input.}

\revise{Data cleaning should ensure data is fit-for-purpose, with the analysis purpose guiding the data quality report to identify relevant data quality issues~\cite{DBLP:journals/jmis/WangS96, DBLP:journals/cacm/PipinoLW02}.} These issues determine data cleaning objectives, which define actionable constraints and operations necessary to improve data quality. Consequently, data cleaning objectives drive the design of data cleaning workflows, as illustrated in Figure~\ref{fig-llma-model}. Despite advancements, data scientists still spend over 80\% of their time on cleaning tasks due to diverse domain requirements~\cite{rezig2019data}. Designing effective workflows often involves multiple rounds of selecting appropriate operations and arguments to ensure accuracy~\cite{stonebraker2018data}. This process remains time-consuming and error-prone, as it demands domain expertise and a deep understanding of the dataset and its schema, yet it is essential for auditing data analyses and ensuring the quality of the resulting tables. 

\revise{Recent advances in large language models (LLMs) have sparked interest in their application to data cleaning~\cite{DBLP:journals/pvldb/EltabakhNAOT24, DBLP:journals/corr/abs-2404-18681, DBLP:journals/corr/abs-2403-08291}. 
However, few existing approaches effectively automate the design of executable data cleaning workflows that are aligned with specific analytical purposes and capable of addressing diverse data quality issues, despite the availability of modern cleaning tools (see Table~\ref{tab:benchmark_lit}). To address this gap, we propose \name\ pipeline and dataset that jointly fulfill three critical criteria: (1) is guided by a data analysis purpose, (2) supports the resolution of multiple data quality issues, and 3) generates executable and reusable data cleaning workflows.}

\begin{table}[t]
\centering
\small
\renewcommand{\arraystretch}{1.05}
\resizebox{\linewidth}{!}{%
\begin{tabular}{l c c c c}
\toprule
\textbf{Frameworks} & \textbf{Task} & \textbf{Workflow} & \textbf{Multi-DQ} & \textbf{Purpose}\\
\midrule
NADEEF~\cite{dallachiesa2013nadeef} & Rule-based DC & \cmark & \cmark &  \\
ActiveClean~\cite{krishnan2016activeclean} & DC for ML &  & \cmark &  \\
Holoclean~\cite{rekatsinas2017holoclean} & ML for DC &  & \cmark &  \\
SAGA~\cite{siddiqi2023saga} & DC for ML & \cmark & \cmark & \cmark*  \\
RetClean~\cite{DBLP:journals/pvldb/EltabakhNAOT24} & LLM for DC &  & \cmark & \\
LLMClean~\cite{DBLP:journals/corr/abs-2404-18681} &LLM for DC &  & \cmark & \\
CleanAgent~\cite{DBLP:journals/corr/abs-2403-08291} &LLM for DC&  &  & \cmark \\
COMET~\cite{mohammed2025step} & DC for ML & \cmark & \cmark &  \\
\midrule
\textbf{AutoDCWorkflow} & LLM for DC
& \cmark & \cmark & \cmark \\
\bottomrule
\end{tabular}
}
\vspace{-5pt}
\caption{
\revise{Comparison of existing data cleaning (DC) frameworks and \textbf{AutoDCWorkflow} in terms of whether the framework (1) generates an executable data cleaning workflow (\textbf{Workflow}), (2) supports multiple data quality issues (\textbf{Multi-DQ}), and (3) is guided by a specific data analysis purpose (\textbf{Purpose}). $*$ refers to the purpose being defined by downstream machine learning performance, rather than a data analysis purpose.}}
\label{tab:benchmark_lit}
\vspace{-5pt}
\end{table}

Two natural questions arise in the context of LLM-assisted data cleaning: (1) Can LLMs help users design efficient, purpose-driven data cleaning workflows? (2) How can we evaluate LLMs' ability to automate these workflows? Therefore, we propose (1) an LLM-based pipeline,~\name, that \textbf{Auto}matically generates a \textbf{D}ata \textbf{C}leaning \textbf{Workflow} given a data analysis purpose, and (2) a new dataset benchmark for evaluating automated workflow generation.

\begin{figure*}[!th] 
\centering 
\includegraphics[width=\linewidth]{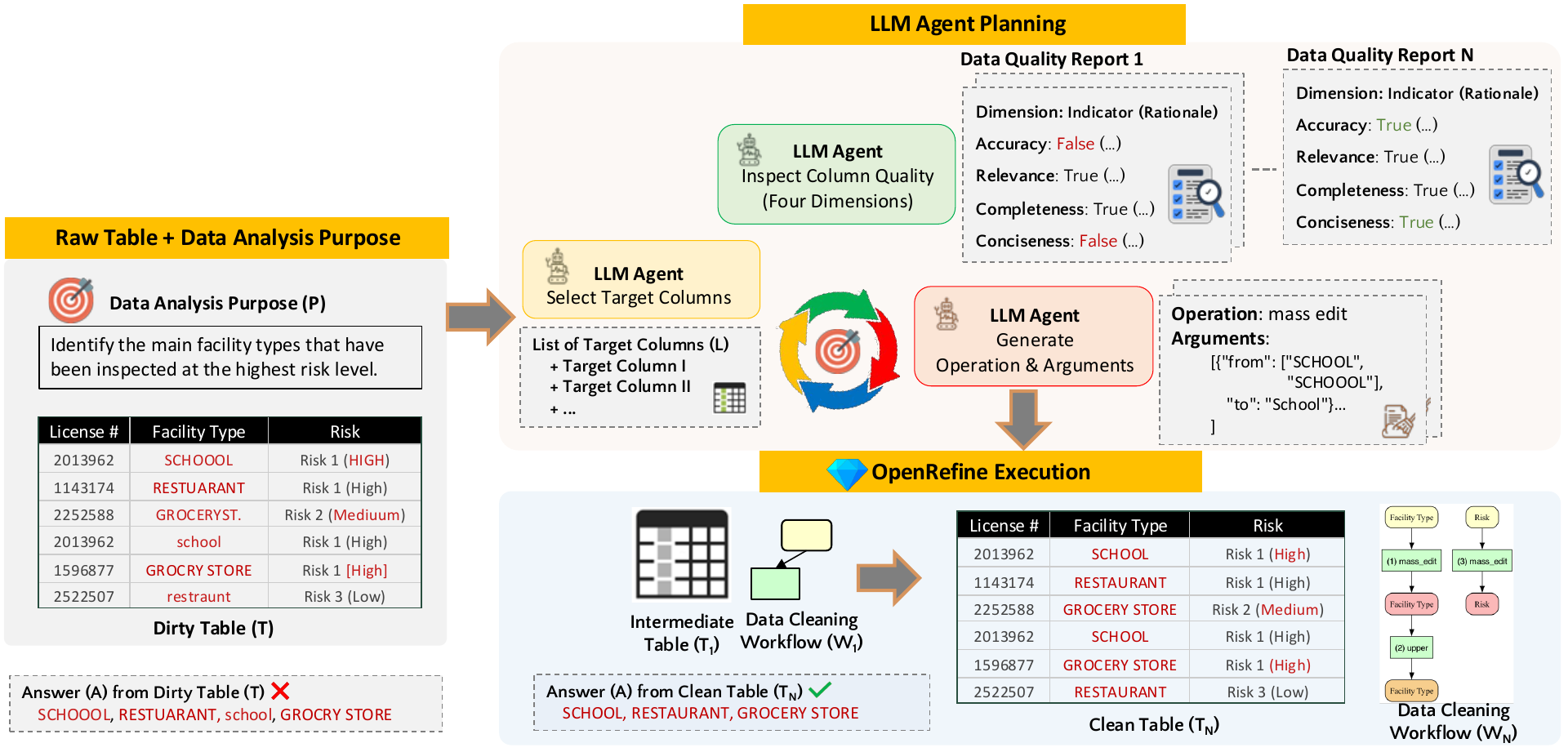} 
\vspace{-20pt}
\caption{Architecture of the \name~framework. \revise{Given a data analysis purpose $\mathbf{P}$: Identify the main facility types that have been inspected at the highest risk level, and a dirty table $\mathbf{T}$, we first run the purpose on $\mathbf{T}$ and obtain an incorrect answer $\mathbf{A}$: ``SCHOOOL, RESTUARANT, school, GROCRY STORE''}. An iterative data cleaning process is then initiated, consisting of: (1) \textit{Selecting the Target Column}, (2) \textit{Inspecting Column Quality}, (3) \textit{Generating Operations \& Arguments}, and (4) \textit{Executing Operations} via the OpenRefine API. \revise{In each iteration, the framework predicts and applies the next operation, evaluates the resulting column quality, and continues until all target columns satisfy the required quality standards (i.e., all dimensions are marked as \texttt{True} in the Data Quality Report). This process yields a cleaned table $\mathbf{T}_N$ and a corresponding complete workflow $\mathbf{W}_N$. Running the purpose on $\mathbf{T}_N$ produces the correct answer $\mathbf{A}$: ``SCHOOL, RESTAURANT, GROCERY STORE''.}}
\label{fig-pipeline}
\end{figure*}

\name~pipeline leverages LLMs as data cleaning agents to assess data quality and generate a sequence of operations based on the analysis purpose and raw table, inspired by the purpose-driven data cleaning model \cite{li2023reusability}. It features three iterative, prompt-based components, as shown in Figure \ref{fig-pipeline}, and continues iterating until no data quality issues remain. Notably, \name~is a companion tool built on OpenRefine. While the pipeline itself is programming language-agnostic, we use OpenRefine operations for demonstration in our experiments.


The benchmark evaluates LLM agents' ability to automate data cleaning workflows. \name~ provides annotated datasets that include purposes, answers, raw tables, manually curated clean tables, and workflows. The raw tables are sampled from six real-world datasets in domains like social science, public health, finance, and transportation. 

\emph{Summary of Contributions.} \name\ introduces a fully automated LLM-driven framework for data cleaning workflow generation. Our benchmark and evaluation metrics establish a foundation for future advancements in LLM-powered data cleaning. Our contributions include:
\begin{itemize}[noitemsep, topsep=0pt]
    \item[(1).] We design a three-stage LLM-based pipeline—Select Target Columns, Inspect Column Quality, and Generate Operations \& Arguments—for automated data cleaning workflow generation.
    \item[(2).] We construct a dataset benchmark to evaluate data cleaning workflows across three dimensions: Purpose Answer, Column Value, and Workflow (Operations).
    \item[(3).] We assess the performance of state-of-the-art models including Llama 3.1–8B, Mistral–7B, \revise{Gemma 2–9B and 27B}. \revise{Gemma 2–9B and 27B stand out, consistently generating plausible workflows that closely resemble human-curated ones, resulting in more accurate purpose-driven answers.}
\end{itemize}

\section{Related Work} \label{sec:related_work}


\noindent\textbf{Data Cleaning for Machine Learning.} 
\revise{Several systems have been developed to improve the quality of training data used in machine learning (ML) pipelines. NADEEF~\cite{dallachiesa2013nadeef} provides a rule-based framework for detecting and repairing data errors using user-defined constraints. ActiveClean~\cite{krishnan2016activeclean} integrates data cleaning with model training by prioritizing cleaning based on the model's sensitivity to dirty data. SAGA~\cite{siddiqi2023saga} and COMET~\cite{mohammed2025step} extend this paradigm by generating executable workflows that address multiple types of data quality issues, optimizing cleaning strategies to maximize downstream model performance. In particular, SAGA advances fit-for-purposes data cleaning by systematically learning and optimizing cleaning strategies based on downstream ML performance. However, these frameworks are primarily designed to enhance model accuracy, rather than being tailored to specific user-defined analytical purposes.}


\noindent\textbf{Machine Learning for Data Cleaning.} \revise{Machine learning techniques have long been used to support data cleaning. Holoclean~\cite{rekatsinas2017holoclean}, for example, leverages probabilistic inference to repair data by modeling integrity constraints and correlations. Pre-trained language models (PLMs) have been applied to entity matching (EM) tasks~\cite{kasai-etal-2019-low, li_deep_2020, suhara2022annotating, fang2023kaer, wadhwa-etal-2024-learning}, focusing primarily on duplicate detection and resolution. 
Recent work has explored the use of large language models (LLMs) to automate various data cleaning tasks. RetClean~\cite{DBLP:journals/pvldb/EltabakhNAOT24} applies retrieval-augmented generation with LLMs to impute missing values and correct erroneous entries. LLMClean~\cite{DBLP:journals/corr/abs-2404-18681} uses LLMs to automatically construct context models—specifically, Ontological Functional Dependencies (OFDs)—to detect and repair inconsistencies in relational data.
CleanAgent~\cite{DBLP:journals/corr/abs-2403-08291} combines LLMs with a declarative API to support standardization tasks such as date formatting and categorical value normalization. While these approaches demonstrate the potential of LLMs in individual cleaning tasks, they do not generate end-to-end, executable workflows tailored to specific analytical purposes.}


\noindent\textbf{LLMs for Tabular Understanding and Transformation.} \revise{
Large language models (LLMs) have been applied to various tabular tasks, including fact verification~\cite{chen2019tabfact} and question answering~\cite{jin2022survey}, and structured reasoning~\cite{Binder}. CHAIN-OF-TABLE~\cite{wang2024chain} demonstrated a step-by-step approach where LLMs simulate reasoning over table transformations via synthetic operations, highlighting potential for structured tabular problem-solving. Gandhi et al.~\cite{gandhi2024better} further showed that clear task descriptions can guide one-shot transformation plans, underscoring the role of instruction design in dataset repurposing.}

\section{Background}

\subsection{Purpose-Driven Data Cleaning Workflows}
\label{sec:problem_definition_purpose}

Our approach is motivated by the principle that ``\textit{high-quality data is fit-for-purpose}", emphasizing that each data cleaning task should be driven by a well-defined data analysis purpose~\cite{paul2010fit, sidi2012data, burden2024fit}. \revise{Accordingly, different purposes may produce distinct result tables and necessitate different data cleaning operations.} The data analysis purposes in our benchmark are summarized in Table~\ref{tab:purpose-category}, categorized and documented as follows:

\noindent\textbf{Descriptive Statistics:} This involves general statistics or aggregates about the dataset. For example, we might ask, ``What’s the highest or lowest loan amount for each zip code?''

\noindent\textbf{Counting and Grouping:} Identifying distinct elements or counting occurrences within the dataset, e.g., ``Count how many types of risks are recorded in the dataset.''

\noindent\textbf{Category or Type Classification:} Classifying data based on certain criteria. For instance, ``Identify which sponsors provide breakfast.''

\noindent\textbf{Time-based Analysis:} Analyzing trends or patterns over time. For example, ``Find dishes that first appeared before the year 2000.''

\noindent\textbf{Correlations and Relationships:} Exploring relationships between different columns in the dataset. For instance, ``Examine if a correlation exists between jobs reported and the loan amount received.''

\noindent\textbf{Filtering and Specific Queries:} Performing specific queries or filtering the dataset based on defined conditions, e.g., ``Report all NAICS Codes that indicate job counts greater than 3.0.''

\begin{table}[t]
\small
\centering
\vspace{-0.45cm}
\resizebox{\columnwidth}{!}{%
\begin{tabular}{c|c}
\hline
\textbf{Category}      & \textbf{Number of Purposes}  \\ \hline
Descriptive Statistics         & 26   \\ \hline
Counting and Grouping        & 27  \\ \hline
Category or Type Classification & 19   \\ \hline
Time-based Analysis             & 17    \\ \hline
Correlations and Relationships  & 10   \\ \hline
Filtering and Specific Queries & 43   \\ \hline
\end{tabular}%
}
\vspace{-5pt}
\caption{Summary of Data Analysis Purposes Categories.}
\label{tab:purpose-category}
\end{table}

\subsection{OpenRefine Workflow for Transparency}

OpenRefine~\cite{OpenRefine} enhances transparency in data cleaning by documenting operations within a workflow, enabling traceability of intermediate tables transformed by each step. The OpenRefine workflow records how the initial dataset $D_0$ evolves into the final version $D_n$ through a sequence of operations $\mathbf{O}_1, \dots, \mathbf{O}_n$ ~\cite{li2021automatic}:

\begin{equation}
  D_0 \stackrel{\mathbf{O}_1}{\leadsto} D_1 \stackrel{\mathbf{O}_2}{\leadsto} D_2 \stackrel{\mathbf{O}_3}{\leadsto} \cdots
  \stackrel{\mathbf{O}_n}{\leadsto}  D_n ~. \label{eq-wf-history}
\end{equation}

A workflow includes (1) data operations $\mathbf{O}_{i}$ applied to transform the data and (2) intermediate tables $D_{i}$ updated at each step. 


\section{Approach Architecture}
\label{sec:arch}

The complete \name~pipeline, shown in Figure \ref{fig-pipeline}, consists of two main stages: (1) LLM agent planning and (2) data cleaning via OpenRefine execution. 

\noindent\revise{\textbf{Problem Definition.} Given a table $\mathbf{T}$ and a data analysis purpose $\mathbf{P}$, \name\ aims to generate data cleaning workflow $\mathbf{W}_N$ and the cleaned table $\mathbf{T}_N$.} To achieve this, \name\ begins by identifying a list of target columns $\mathbf{L}$. Then it iteratively selects and cleans columns and removes them from $\mathbf{L}$ once successfully cleaned.
In the OpenRefine stage, the generated operations and arguments $\mathbf{O}(\cdot)$ are executed to modify the input table into an intermediate table $\mathbf{T}_i$. Each operation $\mathbf{O}_i(\cdot)$ is logged in the data cleaning workflow $\mathbf{W}_i$. The updated table $\mathbf{T}_i$ and workflow $\mathbf{W}_i$ then serve as inputs for the next LLM planning iteration. This process continues until $\mathbf{L}$ is empty, indicating that all target columns have been cleaned in alignment with $\mathbf{P}$. The final cleaned table $\mathbf{T}_N$, produced by the workflow $\mathbf{W}_N$ as $\mathbf{T}_N = \mathbf{W}_N \circ \mathbf{T}$, is claimed to be of high quality and capable of generating the correct answer $\mathbf{A}$ for the purpose $\mathbf{P}$.

\subsection{Select Target Columns}
Purpose-driven data cleaning emphasizes data quality fit-for-purpose~\cite{paul2010fit, sidi2012data, burden2024fit}. Not all columns are necessarily relevant to the purpose of the given analysis. Select Target Columns agent aims to reduce the complexity of data cleaning from the entire table to only the columns pertinent to the given purpose~\cite{gandhi2024better}. 

We design the prompt template with task instructions and few-shot demonstration examples, inspired by~\citet{gandhi2024better}. Each example contains the following content: (1) table information, (2) purpose statement, (3) explanation or rationale for identifying relevant columns based on the purpose statement, and (4) target columns.

\subsection{Inspect Column Quality}

Data quality is a multidimensional concept ~\cite{wand1996anchoring}. \citet{wang2023overview} identify six key dimensions: \textit{completeness}, \textit{accuracy}, \textit{timeliness}, \textit{consistency}, \textit{relevance}, and \textit{concise representation}, \revise{while emphasizing that the choice of dimensions often depends on the study context.}
\revise{In our setting, we exclude \textit{timeliness}, as no timestamps or external knowledge are available to evaluate data freshness, and \textit{consistency}, which is typically defined through rule-based dependencies across multiple columns, is not feasible in OpenRefine's operation model that emphasizes single-column transformations. Accordingly, this paper focuses on four widely adopted dimensions: \textit{accuracy}, \textit{relevance}, \textit{completeness}, and \textit{conciseness}.}


The Inspect Column Quality agent produces a Data Quality Report assessing these four quality dimensions, defined as follows:
\textit{Accuracy} evaluates whether the target column is free from obvious errors, inconsistencies, or biases.
\textit{Relevance} determines if the target column is essential to addressing the analysis objectives.
\textit{Completeness} checks if the target column has an adequate sample size with minimal missing values.
\textit{Conciseness} assesses if spellings are standardized and whether there are no different representations for the same concept.

It is important to note that \textit{relevance} is implicitly assessed by the Select Target Column agent and explicitly reevaluated by the Inspect Column Quality agent. This reevaluation ensures that the values in the selected column remain aligned with the data analysis purpose, thereby reducing the risk of misalignment introduced in the earlier stage.

Data cleaning objectives are specific and focused on addressing errors in the target column. They are generated based on the Data Quality Report if any of the four quality dimensions are evaluated as \textit{False}. This process guides the prediction of the next operation and its arguments.

\subsection{Generate Operation and Arguments}

Instructions and example usages of data cleaning operations serve as prompts to guide \name~in generating appropriate operations and corresponding arguments iteratively. We include six commonly used OpenRefine operations: \texttt{upper}, \texttt{trim}, \texttt{numeric}, \texttt{date}, \texttt{mass\_edit}, and \texttt{regexr\_transform}. The mechanisms of these operations and their associated data quality dimensions are summarized in Table~\ref{table:data-operations}. Crucially, a single operation can impact multiple data quality dimensions. Despite the limited number of operations, the vast space of possible arguments and the flexible combination of operations enable our approach to handle complex real-world data quality issues.

\begin{table*}[t]
\centering
\small
\renewcommand{\arraystretch}{1.05}
\setlength{\tabcolsep}{4pt}
\resizebox{\linewidth}{!}{%
\begin{tabular}{l l l}
\toprule
\textbf{Operation} & \textbf{Description} & \textbf{Quality Dimensions} \\
\midrule
\texttt{upper} & Converts text to uppercase for consistent casing. & Conciseness \\
\texttt{trim} & Removes leading/trailing spaces. & Accuracy, Conciseness \\
\texttt{numeric} & Converts values to numeric format if applicable. & Accuracy \\
\texttt{date} & Standardizes date format across entries. & Accuracy \\
\texttt{mass\_edit} & Merges similar or erroneous values. & Accuracy, Conciseness, Relevance, Completeness \\
\texttt{regexr\_transform} & Uses regex to match and replace patterns. & Accuracy, Conciseness, Relevance \\
\bottomrule
\end{tabular}
}

\caption{Mechanisms and quality dimensions of common data operations.}
\label{table:data-operations}
\end{table*}

\emph{Operation Generation.} We design a structured prompt template  (see Appendix~\ref{sec:appendix_eod}). This template incorporates task instructions, operation definitions, examples, and explanations. The LLM agent determines the operation that best fits the given table contents and data analysis purpose.

\emph{Arguments Generation.} The \texttt{mass\_edit} and \texttt{regexr\_transform} operations require argument generation to handle various data errors (see Appendix~\ref{sec:appendix_ops}). We provide few-shot examples demonstrating arguments usage for specific data cleaning objectives.

\begin{table}[!htb]
    \centering
    \resizebox{\columnwidth}{!}{%
    \begin{tabular}{l|c|c|c|c|c|c}
    \hline
    \textbf{Dataset} & \textbf{\# Purposes}  & \textbf{\# Answers} &\textbf{\# Tables}  & \textbf{\# Workflows} & \textbf{\# Rows}&
    \textbf{\# Columns}\\ \hline
    Menu     &30   &30  &3    &30 & 100&20\\ \hline
    Dish     &16   &16  &9    &16 & 50&8\\ \hline
    Flights  &16   &16  &16   &16 & 50&7\\ \hline
    PPP      &22   &22  &11   &22 & 20&14\\ \hline
    Hospital &28   &28  &28   &28 & 20&12\\ \hline 
    CFI      &30   &30  &29   &30 & 10&11\\ \hline\hline
    Total    &142  &142 &96   &142 & -&-\\ \hline
    \end{tabular}%
    }
\caption{\revise{Summary of Benchmark Dataset Statistics.  Total integrates six datasets. Each purpose is assigned to a single table, and the \# Tables represents the number of distinct tables in each dataset.}}
\label{tab:benchmark_stats}
\end{table}

\section{Benchmark Demonstration}
The benchmark includes annotated datasets and evaluation dimensions to assess workflow quality. The benchmark dataset comprises data analysis purposes, answer sets, raw tables\footnote{The number of tables does not match the number of purposes in Menu, Dish, PPP, and CFI because these datasets already contain inherent errors.}, and manually curated tables and workflows. The statistic of the benchmark dataset is provided in Table~\ref{tab:benchmark_stats}. The benchmark evaluation enables a robust assessment of generated workflows in three key dimensions:\\
\textbf{[RQ1]} Purpose Answer Dimension: Can we retrieve approximately or exactly the same correct answer from the repaired clean table?\\
\textbf{[RQ2]} Column Value Dimension: How closely does the repaired table resemble the human-curated table?\\
\textbf{[RQ3]} Workflow (Operation) Dimension: To what extent does \name~generate the correct and complete set of operations?

\subsection{Dataset Construction}

\subsubsection{Dataset Preparation and Partition} \label{sec:bench_dataset}
\revise{The benchmark spans multiple domains, including Menu and Dish\footnote{\url{https://menus.nypl.org/}} (social science), Chicago Food Inspection (CFI)\footnote{\url{https://data.cityofchicago.org/Health-Human-Services/Food-Inspections/}} and Hospital (public health)~\cite{mahdavi2019reds}, Paycheck Protection Program (PPP)\footnote{\url{https://data.sba.gov/dataset/ppp-foia/}} (finance), and Flights (transportation)~\cite{mahdavi2019reds}. To evaluate the LLM's ability to generate workflows on datasets with diverse data types and varying levels of data quality issues, we sample tables from these original datasets and systematically inject different types of errors based on their associated analysis purposes.}

\revise{We implement Python functions to inject common real-world data quality issues into clean datasets. These include: (1). Duplicate variants: Slightly different but semantically equivalent values (e.g., "McDonalds" vs. "Mc Donald’s"). (2). Formatting inconsistencies: Extra whitespace and non-ASCII characters. (3). Case variations: Mixed use of uppercase and lowercase letters. (4). Type errors: Inserting strings into numeric fields (e.g., "N/A" in a price column). These errors simulate real-world data quality challenges, allowing us to assess the model’s robustness in data cleaning. The error injection rate can be found in Table~\ref{tab:dataset-error-ratio}.} Moreover, empty columns are removed, as they do not contribute meaningful information for answering the given purposes. 

\subsubsection{Purpose and Answer Annotation}

\name~benchmark defines 142 purposes based on the sample rows and column schema \revise{(see Table~\ref{tab:purpose_stats})}. These purposes are constructed using two methods: referencing columns by exact or paraphrased names and specifying specific data values. The first method allows LLMs to recognize and identify target columns, while the second adds complexity by requiring inference of relationships between cell values and the column schema.

To assess the difficulty of each purpose, we consider two factors: (1) the number of target columns involved and (2) the diversity of their data types. We hypothesize that purposes requiring more target columns and diverse data types provide a more challenging task for LLMs to generate workflow. \revise{For instance}, PPP tables primarily contain numerical data with less variation than other datasets, making them a relatively simpler task for LLMs.


\begin{table}[!ht]
\centering
\small
\renewcommand{\arraystretch}{1.1}
\setlength{\tabcolsep}{6pt}
\begin{tabular}{lcc}
\hline
\textbf{Dataset} & \textbf{\# Tables} & \textbf{Error Ratio (mean ± std)} \\
\midrule
Menu     & 30 & 0.0636 ± 0.1078 \\
Dish     & 16 & 0.1480 ± 0.0745 \\
Flights  & 16 & 0.2584 ± 0.0942 \\
PPP      & 22 & 0.3451 ± 0.1930 \\
Hospital & 28 & 0.1463 ± 0.0498 \\
CFI      & 30 & 0.2307 ± 0.0545 \\
\bottomrule
\end{tabular}
\caption{Error injection rates across tables.}
\label{tab:dataset-error-ratio}
\end{table}

\emph{Answer Annotation.} Querying the dataset returns purpose answers in three forms: a single numeric value, a string, or a collection of columns and cell values. This enables us to compare the purpose answers derived from LLM-generated tables with those from ground truth tables.

\begin{table}[!hbt]
\centering
\resizebox{\columnwidth}{!}{%
\begin{tabular}{c|c|ccc|ccc}
\hline
\multirow{2}{*}{Dataset} &
  \multirow{2}{*}{\# Purposes} &
  \multicolumn{3}{c|}{\textbf{\# Target Columns}} &
  \multicolumn{3}{c}{\textbf{Target Column Types}} \\ \cline{3-8} 
 &
   &
  \multicolumn{1}{c|}{$\mathbf{1}$} &
  \multicolumn{
  1}{c|}{$\mathbf{2}$} &
  $\mathbf{\geq 3}$ &
  \multicolumn{1}{c|}{\textbf{Numeric}} &
  \multicolumn{1}{c|}{\textbf{Text}} &
  \textbf{Date} \\ \hline
Menu    & 30 & \multicolumn{1}{c|}{16} & \multicolumn{1}{c|}{13}  & 1 & \multicolumn{1}{c|}{11}  & \multicolumn{1}{c|}{23} & 2\\ \hline
Dish    & 16 & \multicolumn{1}{c|}{1} & \multicolumn{1}{c|}{6}  & 9 & \multicolumn{1}{c|}{11} & \multicolumn{1}{c|}{15} & 5 \\ \hline
Flights    & 16 & \multicolumn{1}{c|}{4} & \multicolumn{1}{c|}{8}  & 4 & \multicolumn{1}{c|}{0} & \multicolumn{1}{c|}{6} &  13\\ \hline
PPP     & 22 & \multicolumn{1}{c|}{4} & \multicolumn{1}{c|}{16} & 2 & \multicolumn{1}{c|}{21} & \multicolumn{1}{c|}{13} & 0 \\ \hline
Hospital     & 28 & \multicolumn{1}{c|}{3} & \multicolumn{1}{c|}{15} & 10 & \multicolumn{1}{c|}{6} & \multicolumn{1}{c|}{28} & 0 \\ \hline
CFI & 30 & \multicolumn{1}{c|}{10} & \multicolumn{1}{c|}{16}  & 4 & \multicolumn{1}{c|}{4}  & \multicolumn{1}{c|}{29} &  4\\ \hline

\end{tabular}%
}
\caption{Summary of Data Analysis Purposes Statistics. Each row represents the number of target columns and column types per purpose, providing insights into the complexity of data analysis purposes across different datasets.}
\label{tab:purpose_stats}

\end{table}

\subsubsection{Workflow Annotation}

Different human curators may prefer different data cleaning operations, leading to varied workflows for the same data cleaning task. As a result, there is no single gold-standard workflow. \revise{In this study, a data cleaning expert utilized OpenRefine operations, cleaning the given table for a data analysis purpose and annotated data cleaning workflows as a reference, which we regard as \textit{silver-standard} ground truth for evaluating LLM-generated workflows.} We assess recall and accuracy by comparing the operations predicted by LLMs to those annotated in the \textit{silver} workflows. This comparison assesses how closely LLMs approximate the silver-standard ground truth rather than serve as a strict indicator of workflow correctness. 

\begin{figure}[!hbt] 
\centering 
\includegraphics[width=1.02\linewidth]{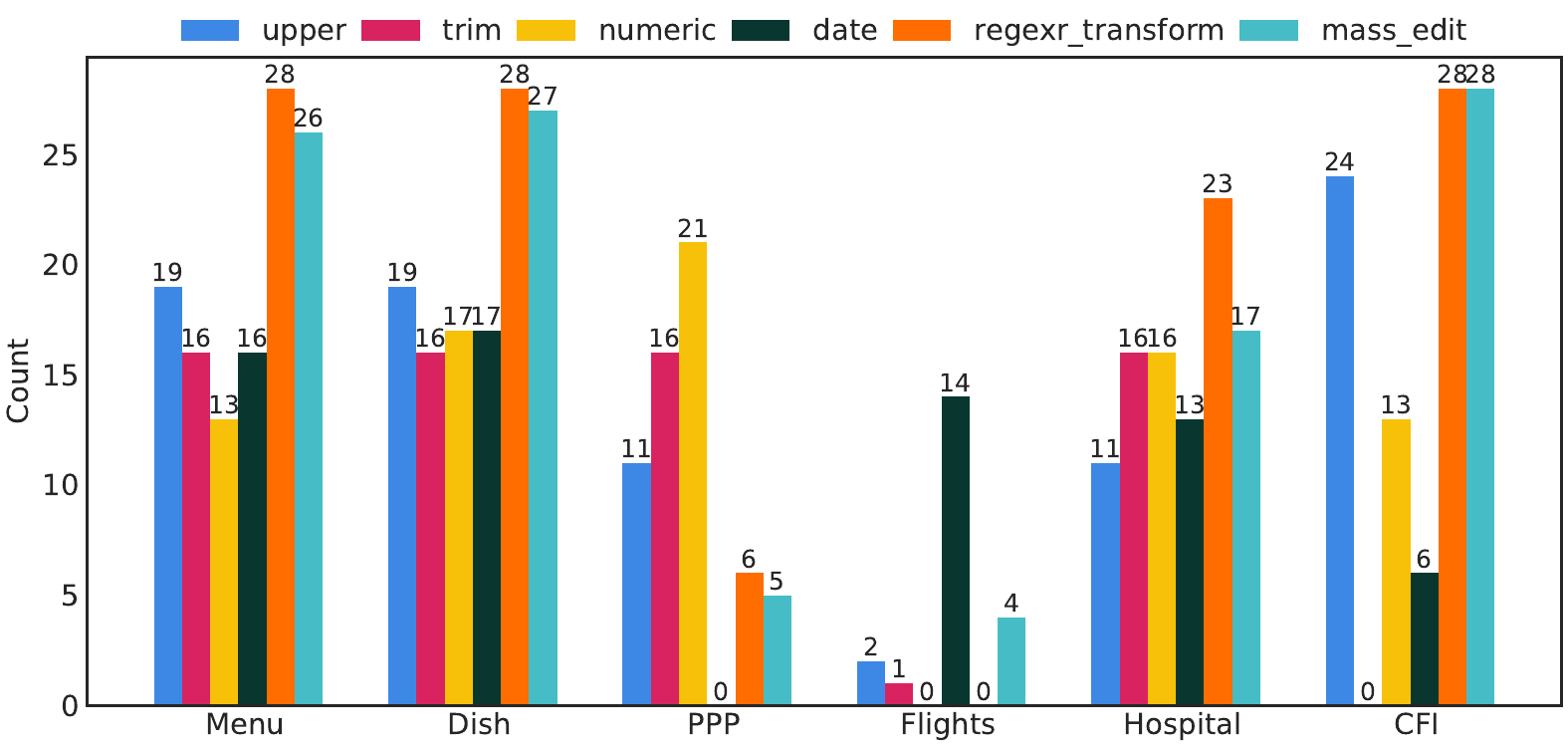} 

\caption{Frequency distribution of data operation types across all datasets.}
\label{fig-purposes-stats}
\end{figure}

Since datasets differ in data types and the distribution of quality issues, each dataset has its own most frequently used operations. For instance, \textit{numeric} is the most frequently used operation in PPP, while \textit{date} is the most frequently used in Flights, as shown in Figure~\ref{fig-purposes-stats}.

\subsection{Evaluation Dimensions and Metrics}

We assess the generated workflows across three key dimensions: (1) \textbf{Accuracy}, by evaluating whether the cleaned table produces the correct answer for the given purpose (RQ1 Answer Dimension); (2) \textbf{Consistency}, by measuring how closely the repaired table aligns with the ground truth (RQ2 Column Value Dimension); and (3) \textbf{Plausibility}, determining the extent to which the generated workflow resembles the human-curated ground truth (RQ3 Workflow Dimension).

\subsubsection{Purpose Answer Dimension}

The purpose answer dimension assesses the accuracy of the workflow by measuring how well the cleaned table produces correct answers aligned with the ground truth. We evaluate alignment using \textit{Precision}, \textit{Recall}, \textit{F1}, and \textit{Similarity}. Exact matches are used for single-valued answers (e.g., numbers or strings), while complex answers (e.g., lists or tables in JSON format) are assessed element-wise with \textit{Precision}, \textit{Recall}, and \textit{F1} scores. 
The \textit{Similarity} quantifies the proportion of matching characters relative to the total number of characters across both sequences. This ensures a comprehensive evaluation of workflow accuracy across different answer types and complexities. \revise{The matching characters are captured by finding the longest contiguous matching block, recursively applying the same logic to the unmatched regions to the left and right of the match, and summing all matching blocks to compute total matches. }



\subsubsection{Column Value Dimension} \label{eva-column-dimension}

This dimension assesses the consistency of data cleaning workflows by measuring how closely cleaned column values align with the ground truth table. The column cleanness ratio represents the average proportion of matching values across all target columns. Equivalence comparison is performed based on data types: numeric, string, and date. The detailed formula for calculating the column cleanness ratio \revise{is shown in Equation~\ref{eq:average_match_ratio}.}
\begin{equation} 
\label{eq:average_match_ratio}
\text{Ratio} = \frac{1}{M} \sum_{j=1}^{M} \frac{1}{N} \sum_{i=1}^{N} \delta(T_{i,j}, G_{i,j})
\end{equation}

\begin{small}
\noindent
\text{where: } 
$M$: Total number of target columns, 
$N$: Total number of rows, 
$T_{i,j}$: Value in row $i$ and column $j$ of the table, 
$G_{i,j}$: Value in row $i$ and column $j$ of the ground truth, \\
$\delta(T_{i,j}, G_{i,j}):$ 
\[
\begin{cases}
1, & \text{if } T_{i,j} \text{ is numeric or } \text{lower}(T_{i,j}) = \text{lower}(G_{i,j}) \\
0, & \text{otherwise}
\end{cases}
\]
\end{small}


\begin{table*}[t]
\centering
\small
\renewcommand{\arraystretch}{1.1}
\setlength{\tabcolsep}{4.5pt}
\resizebox{\textwidth}{!}{%
\begin{tabular}{lcccc|c|cccc}
\toprule
& \multicolumn{4}{c|}{\textbf{Answer Dimension}} 
& \textbf{Column Dimension} 
& \multicolumn{4}{c}{\textbf{Workflow Dimension}} \\
\cmidrule(lr){2-5} \cmidrule(lr){6-6} \cmidrule(lr){7-10}
\textbf{Model} &
\textbf{Precision} & \textbf{Recall} & \textbf{F1} & \textbf{Similarity} &
\textbf{Ratio} &
\textbf{Exact} & \textbf{Precision} & \textbf{Recall} & \textbf{F1} \\
\midrule
Baseline (Raw Tables) & 0.2345 & 0.2375 & 0.2201 & 0.4678 & 0.4262 & -- & -- & -- & -- \\
\midrule
Llama 3.1–8B (DP) & 0.3070 & 0.2914 & 0.2811 & 0.5039 & 0.5817 & 0.0493 & 0.3768 & 0.1554 & 0.2053 \\
Llama 3.1–8B (\name) & 0.5415{\scriptsize **++} & 0.5185{\scriptsize **++} & 0.5181{\scriptsize **++} & 0.6868{\scriptsize **++} & 0.7475{\scriptsize **++} & 0.0986 & \textbf{0.9331}{\scriptsize ++} & 0.5265 {\scriptsize ++} & 0.6447{\scriptsize ++} \\
\midrule
Mistral–7B (DP) & 0.2862 & 0.2740 & 0.2612 & 0.4933 & 0.4884 & 0.0000 & 0.1162 & 0.0363 & 0.0536 \\
Mistral-7B (\name) & 0.3635{\scriptsize **} & 0.3320{\scriptsize*} & 0.3320{\scriptsize*} & 0.5208 & 0.6004{\scriptsize **++} & 0.0493{\scriptsize ++} & 0.7656{\scriptsize ++} & 0.4096{\scriptsize ++} & 0.5065{\scriptsize ++} \\
\midrule
Gemma2–9B (DP) & 0.3368 & 0.3098 & 0.3030 & 0.5109 & 0.6032 & 0.0141 & 0.4131 & 0.1438 & 0.2042 \\
Gemma2–9B (\name) & 0.3841{\scriptsize **++} & 0.3502{\scriptsize **+} & 0.3479{\scriptsize **++} & 0.5543{\scriptsize +} & 0.6575{\scriptsize **++} & \textbf{0.1056}{\scriptsize++} & 0.8853 {\scriptsize++} & \textbf{0.5698}{\scriptsize++} & \textbf{0.6640}{\scriptsize ++} \\
\midrule
Gemma2–27B (DP) & 0.3298 & 0.3039 & 0.2956 & 0.5073 & 0.5967 & 0.0141 & 0.2887 & 0.0937 & 0.1353 \\
Gemma2–27B (\name) & \textbf{0.6590}{\scriptsize **++} & \textbf{0.6234}{\scriptsize **++} & \textbf{0.6256}{\scriptsize **++} & \textbf{0.7567}{\scriptsize **++} & \textbf{0.7900}{\scriptsize **++} & 0.0775{\scriptsize +} & 0.8759{\scriptsize ++} & 0.5514{\scriptsize ++} & 0.6456{\scriptsize ++} \\
\bottomrule
\end{tabular}
}
\caption{\revise{Overall performance on the combined dataset, including PPP, CFI, Dish, Menu, Hospital, and Flights. Baseline (Raw Table) uses raw tables without cleaning. DP refers to directly using a single prompt for LLMs to generate appropriate data cleaning operations. \textbf{Bold} indicates the highest result in each column. \name\ results are compared with baseline (*) and DP (+) with statistical paired t-test. \textbf{**}(*) represents a 99\% (95\%) confidence level compared with baseline (Raw Table). \textbf{++}(+) represent a 99\% (95\%) confidence level compared with direct prompting (DP).}}
\label{tab:total_performance}
\end{table*}

\subsubsection{Workflow (Operation) Dimension}

To assess the plausibility of generated workflows, this dimension evaluates how well LLM-generated operations align with \revise{the correct operations in the} \textit{silver} workflows. We quantify this alignment using \textit{Precision}, \textit{Recall}, and \textit{F1-score}, which measure the degree of overlap between operations in both workflows. Discrepancies between LLM-generated and \textit{silver} workflows highlight how models differ in the selection of operations resolving the same data quality issues. These variations reflect the model's ability to apply different yet valid approaches to achieve the same data cleaning objective, showcasing distinct ``flavor'' of workflows while still fulfilling the given purpose.

\section{Experiments} \label{sec:experiment}

\subsection{Experiment Settings}

We use raw table inputs (without any prior data cleaning) and a direct prompting (DP) approach as the baseline. In the DP setting, the LLM is instructed to generate the entire data cleaning workflow in a single step. Unlike the iterative pipeline in \name, DP receives the same context—including the list of available operations and example workflows—but lacks access to intermediate reasoning stages and multi-turn decision-making. We evaluate \name\ using three widely adopted, open-access LLMs of comparable model sizes: Llama 3.1-8B~\cite{DBLP:journals/corr/abs-2407-21783}, Gemma 2-9B~\cite{DBLP:journals/corr/abs-2408-00118}, and Mistral-7B~\cite{DBLP:journals/corr/abs-2310-06825}. To assess the impact of model size, we additionally include Gemma 2-27B~\cite{DBLP:journals/corr/abs-2408-00118} in our comparison. This allows us to investigate how scaling model size influences workflow generation. For operation prediction, the models are configured with a default \emph{temperature} of 0.1, which is increased to 0.3 if no valid operation is returned (see more hyperparameter settings in Appendix~\ref{appendix:exp_hyperparams}). \revise{The total number of tokens for all input prompts, excluding table data, is approximately 4538 tokens. The maximum number of output tokens is also set to 2048, with the stop word defined as $[``\backslash n\backslash n\backslash n"]$.} The experiment is conducted using four Nvidia A100 GPUs. Our implementation code is available in an anonymous repository\footnote{\url{https://github.com/LanLi2017/LLM4DC}}.

\subsection{Experiment Results}
\revise{We compare AutoDCWorkflow against two baselines: (1) raw tables, and (2) Direct Prompting (DP). As shown in Table~\ref{tab:total_performance}, AutoDCWorkflow outperforms both baselines across all evaluation dimensions.} A detailed performance table of six datasets can be found in Appendix~\ref{appendix:exp_results}.

\noindent\textbf{Comparison with Baseline for Comparable Models.}
\revise{In the Answer Dimension, AutoDCWorkflow consistently outperforms both baselines in precision, recall, and F1. Notably, Llama 3.1–8B and Gemma 2–9B achieve statistically significant improvements in F1 over both baselines at the 99\% confidence level. For instance, Llama 3.1-8B improves from 0.2201 (Raw Table) and 0.2811 (DP) to 0.5181, and Gemma 2–9B from 0.3030 (DP) to 0.3479.}

\revise{In the Column Dimension, all models using AutoDCWorkflow significantly improve the column value matching ratio compared to both baselines. For example, Llama 3.1–8B rises from 0.4262 (raw) and 0.5817 (DP) to 0.7475, while Gemma 2-9B reaches 0.6575. Even Mistral–7B, the smallest model, shows a strong gain (from 0.4884 to 0.6004), all statistically significant at the 99\% level.}

\revise{In the Workflow Dimension, AutoDCWorkflow substantially improves precision, recall, and F1 across all models relative to DP, again with 99\% confidence. These results highlight the effectiveness of our iterative pipeline in producing higher-quality, purpose-aligned workflows, addressing the limitations of direct prompting.}

\noindent\textbf{Performance of Large-Scale Model (Gemma 2–27B).}
\revise{Among all models, Gemma 2–27B with AutoDCWorkflow achieves the best performance across nearly every dimension. It reaches the highest F1 score in the Answer Dimension (0.6256), the best column matching ratio (0.7900), and strong workflow metrics (precision 0.8759, F1 0.6456), all statistically significant at the 99\% confidence level. These results suggest that the larger model (27B) benefits more from the iterative, purpose-driven cleaning process. This finding indicates exploring a potential scaling law in data cleaning workflow generation, whether increasing model size leads to better reasoning, quality inspection, and workflow generation. Investigating this relationship would be a promising direction for future work.}

\begin{figure}[!ht] 
\centering 
\includegraphics[width=\linewidth]{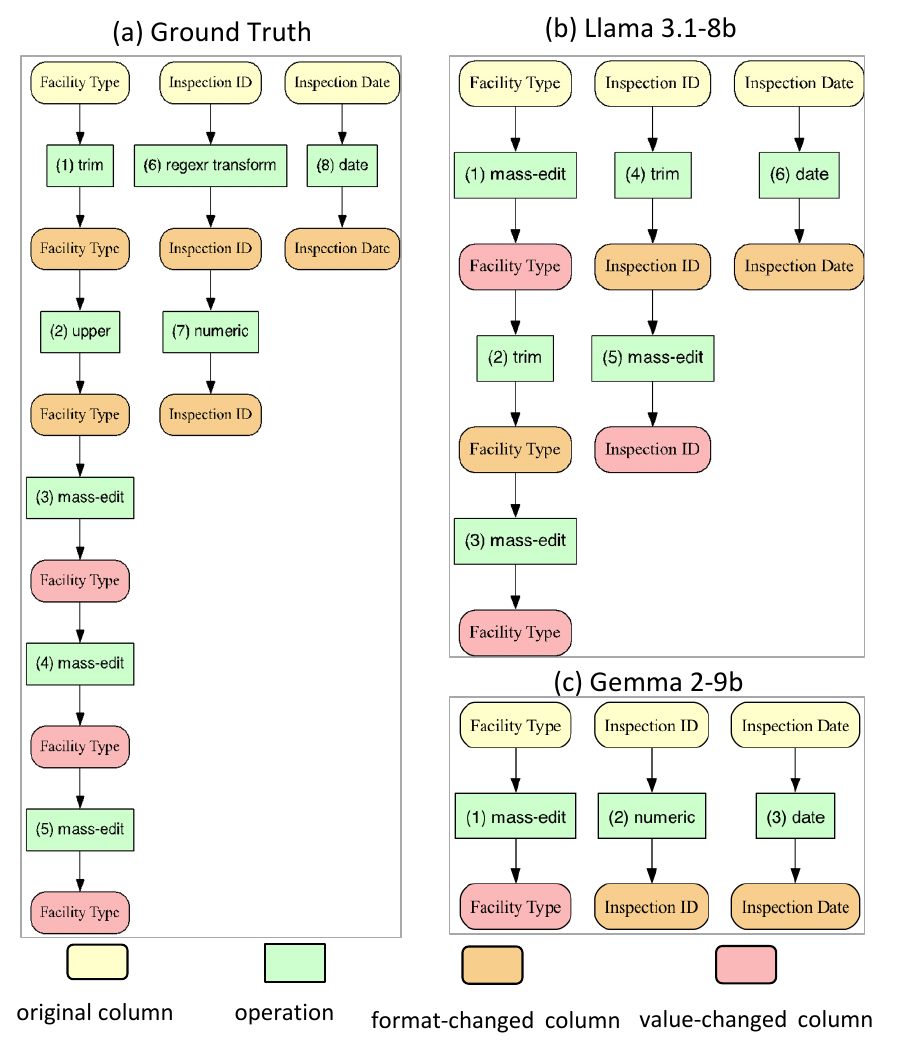}
\caption{\revise{Three workflow versions  generated by (a) Ground Truth (human curators), (b) Llama 3.1-8B, and (c) Gemma 2-9B.}}
\vspace{-8pt}
\label{fig:wf_orma}
\end{figure}

\section{Case Study} \label{sec:case_study}

In this section, we further analyze how different models generate varying workflows by examining their outputs. \revise{We showcase how Llama 3.1-8B and Gemma 2-9B construct workflows with different operations while successfully repairing data to produce the correct answer.} The analysis focuses on the specific purpose of identifying the facility type inspected multiple times on the most recent date.

Figure~\ref{fig:wf_orma} visualizes the workflows using ORMA\footnote{Arrows indicate the flow of data cleaning operations across specific columns.}, where each workflow is modeled as a graph with \textit{data nodes} (colorful rounded rectangles representing columns) and \textit{operation nodes} (green boxes representing applied operations) \cite{li2021automatic}. The three branches in each diagram correspond to the target columns—Facility Type, Inspection ID, and Inspection Date—allowing a direct comparison of how each model processes them.

\revise{Facility Type Column: The ground truth workflow applies \textit{trim} (1x), \textit{upper} (1x), and \textit{mass\_edit} (3x). Llama 3.1-8b correctly selects \textit{trim} (1x) and \textit{mass\_edit} (2x), while Gemma 2-9b consolidates all repairs into a single \textit{mass\_edit} operation. Inspection ID Column: The ground truth workflow uses \textit{regexr\_transform} and \textit{numeric} to standardize non-numeric IDs. Llama 3.1-8B uses \textit{mass\_edit} to accomplish the same result. Inspection Date Column: Both models detect date-time inconsistencies and apply the \textit{date} operation for correction. Despite using fewer steps and varying approaches, both Llama 3.1-8B and Gemma 2-9B effectively clean the table and produce the correct purpose answer: "RESTAURANT". This demonstrates their ability to generate accurate workflows efficiently.}

\section{Conclusions}
This paper explores Large Language Models (LLMs) as automated data cleaning agents for generating fit-for-purpose workflows.  We introduce \name, an iterative framework with three LLM-driven components: selecting target columns, inspecting data quality, and generating operations with arguments. To our knowledge, this is the first fully automated approach to data cleaning workflow generation. To evaluate its effectiveness, we develop a benchmark dataset and tailored evaluation methods, assessing LLMs' ability to infer and apply necessary cleaning steps across diverse datasets. \revise{Experimental results show that all models improve data quality, with Gemma 2-27B producing the most accurate tables and answers, and Gemma 2-9B generating workflows closest to human annotations. This work highlights the potential of LLMs to automate complex data preparation tasks while promoting transparency, reusability, and purpose alignment.}



\section{Limitations}
\label{sec:limitation}
In this section, we introduce the limitations and aim to draw interest in LLMs for data cleaning tasks.

Limitation I: Handling High Data Noise
When the data is highly noisy, LLMs may struggle to learn from the provided examples. Instead of simply adding more in-context learning examples, an alternative approach is to incorporate external knowledge, such as metadata and domain-specific information, to help LLMs better understand the correct data format and improve repair accuracy. One potential direction is self-evolving prompting, where demonstration examples are dynamically refined based on previous model outputs and performance, enabling LLMs to improve their data cleaning capabilities iteratively.

Limitation II: Error Propagation.
In our fully automated framework, errors from incorrect operations can propagate through iterations, compounding their impact on the final table. Since the process lacks human intervention, a single mistake in an early step can cascade through subsequent operations, further influencing the data quality. Future work could introduce human-in-the-loop interactions or feedback mechanisms for LLM agents to detect and correct errors dynamically, preventing their accumulation.

Limitation III: Single-column Repairing. \name\ processes tables column by column independently, ignoring potential correlations and functional dependencies between target columns. This may limit contextual evidence for LLMs, affecting the accuracy and effectiveness of the generated operations. Future work could explore multi-column reasoning to improve operations coherence and consistency. \revise{Adding multi-column operations would enable the framework to handle more complex data cleaning issues, including those that involve relationships between columns. This would be an exciting potential research direction for follow-up studies.}

Limitation IV: Computational resources. Due to the limitation of computational resources, the largest LLM attempted is Gemma 2-27B.

\revise{Limitation V: Sampling Bias. Sampling a subset of column values could miss certain error types, especially rare ones. To mitigate this, our current system performs iterative sampling for larger columns (e.g., with hundreds of instances). We continue sampling until the LLM-generated data quality report rates all four dimensions as "good" for multiple rounds, which serves as a heuristic proxy for coverage. Adaptive sampling and LLM-guided profiling that dynamically adjusts the sample based on initial LLM responses or uncertainty estimates could help increase the efficiency. Another sampling bias would happen for demonstration examples. Our pipeline uses in-context learning with instructions and few-shot examples to inject prior knowledge into the prompts, focusing on general-purpose LLMs without fine-tuning. While adapting to highly specialized domains may require additional techniques such as fine-tuning or retrieval-augmented generation (RAG). These are beyond our current scope and present a promising direction for future work.}

\bibliography{ref}

\appendix

\section{Appendix}

\subsection{OpenRefine Workflow Demonstration}
\label{sec:openrefine-wf-demo}
\revise{Unlike script-based data cleaning, OpenRefine offers a more transparent approach by explicitly documenting data cleaning operations within a workflow, enabling traceability of intermediate tables transformed by each operation.}

\begin{figure}[!bth] 
\centering 
\includegraphics[width=\columnwidth]{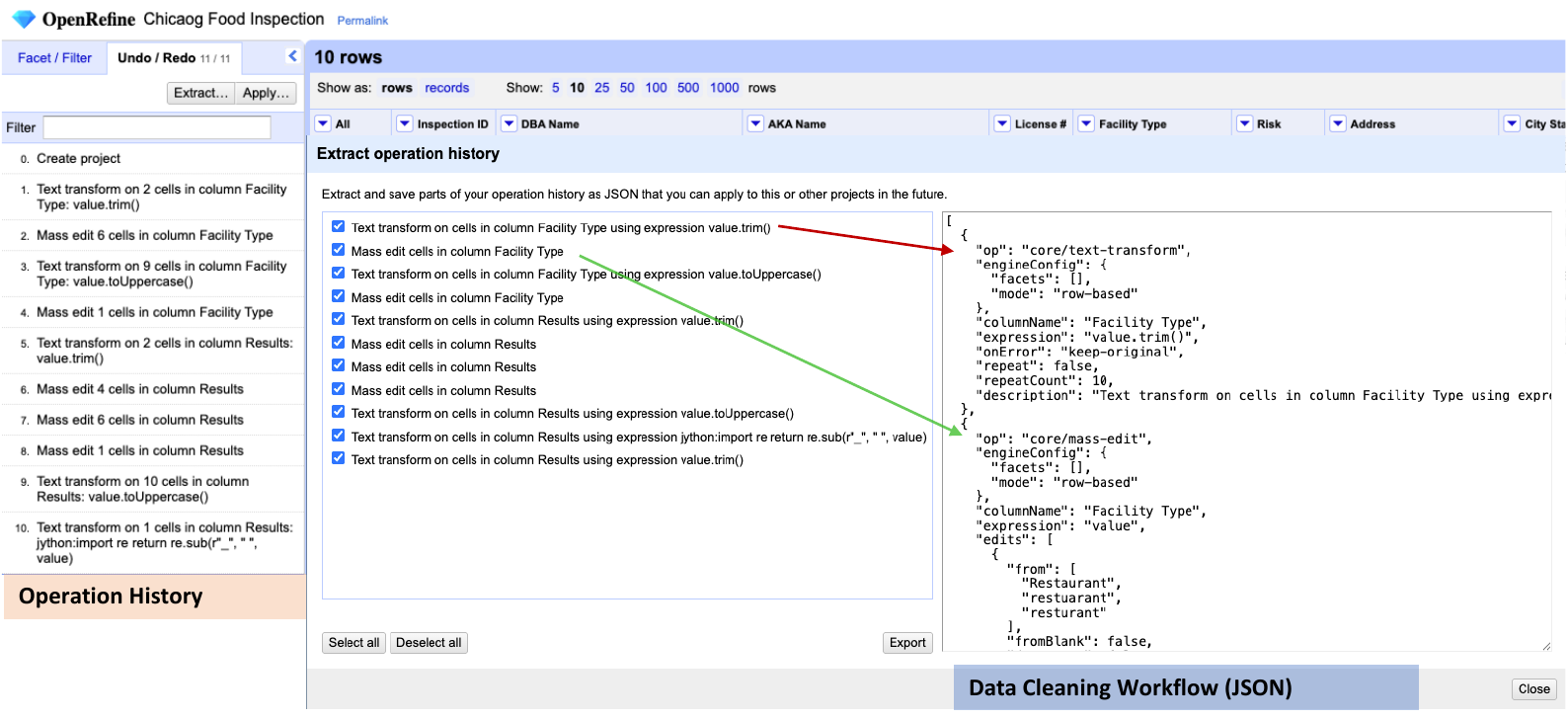} 
\caption{\revise{As users interactively manipulate the dataset, OpenRefine automatically records each operation into the \emph{Operation History} panel (left). These operations can then be exported as a structured \emph{Data Cleaning Workflow} in JSON format (right), enabling reproducibility and reuse across projects.}}
\vspace{-5pt}
\label{fig-or-recipe}
\end{figure}

\revise{OpenRefine records the workflow in a JSON format file, which includes detailed information about each operation, such as the operation name, functions, and other relevant parameters (see Figure \ref{fig-or-recipe}). This workflow provides a clear and reproducible record of how the initial dataset is transformed into its final version through a sequence of data cleaning operations.}

\subsection{Dataset Description}

We include six datasets across six topics in our benchmark: Menu, Dish, Chicago Food Inspection (CFI) data, Paycheck Protection Program (PPP) loan data, Hospital and Flights datasets.

Every row of the Menu table represents a menu record, capturing both ``external'' and ``internal'' information. For instance, the column \textit{event} records the ``external'' information regarding the meal types associated with the menu, such as \textit{dinner}, \textit{lunch}, {breakfast} or other special events. Meanwhile, the column \textit{page\_count} represents the ``internal'' information, documenting the number of pages for this menu. Records in the Dish table capture details about dishes featured on menus, including dish names, frequency of appearance, first and last years on the menu, and highest and lowest prices. This information helps analyze trends in dish popularity over time.

For every row in the CFI dataset sample tables, the data represents inspection results for restaurants and food facilities across Illinois. The column schema contains both inspection information and establishment information. Inspection-related columns include: \textit{Inspection ID}, \textit{Risk}, \textit{Inspection Date}, \textit{Inspection Type} and \textit{Results}. Establishment details are recorded in columns: \textit{DBA Name} (the legal name of the establishment), \textit{AKA Name} (public alias or commonly known name), \textit{License Number}, \textit{Facility Type}, \textit{Street Address}, \textit{City}, \textit{State} and \textit{Zip Code} of the facility. 

Each row in the PPP dataset describes loan information, with columns detailing the loan amount, business type, lender names, race and ethnicity, gender, and location information of the company.

The Hospital and Flights datasets are sourced from previous work \cite{mahdavi2019reds}. The Hospital dataset serves as a benchmark dataset in the data cleaning literature \cite{chu2013holistic, dallachiesa2013nadeef}. It originates from the Hospital Compare website\footnote{http://hospitalcompare.hhs.gov}, maintained by the U.S. Department of Health and Human Services. The website provides data on the quality of care at over 4,000 Medicare-certified hospitals across the United States. The dataset includes key attributes such as provider number, hospital name, address, city, state, zip code, county name, phone number, hospital type, hospital ownership, and availability of emergency services.

The Flights dataset contains information about airline carriers, including details about their flights, scheduled departure and arrival times, and actual departure and arrival times. This dataset provides insights into flight schedules and potential delays by comparing planned and actual flight timings.

\subsection{Mass\_Edit \& Regexr\_transform:  Arguments Introduction}
\label{sec:appendix_ops}

For \textit{mass\_edit}, a list of dictionaries must be created, where each dictionary defines value replacements. In each dictionary, the \emph{key-value} pairs represent groups of similar data instances that need to be standardized to a single, correctly formatted target value. Noted that in OpenRefine, this operation is completed in multi-steps: (1). Data curators select clustering functions and parameters. (2). OpenRefine applies the clustering functions, groups similar values, and suggests a target value for standardization. (3). Data curators review the clustering results and confirm the target values. (4). The final value replacements are then applied. With LLMs, this operation can leverage the model's domain knowledge to recognize similar data instances, enabling the completion of \textit{mass\_edit} without manual verification.

On the other hand, the \textit{regexr\_transform }operation requires an ``ad-hoc" Python-like code implementation as its argument. For example, the cell values in the target column \emph{Year}—such as \emph{Feyerabend,1975,} \emph{Collins,1985}, and \emph{Stanford,2006}--must be transformed to address the purpose of listing all published year information for the respective books. The example function (\ref{eq:regex_example}) can be explained as follows:
(1). ``jython"  signals that the following code is Python code. (2). The ``return"  statement is used to end the execution of the function. (3). The code between "jython:" and ``return" parses and captures patterns in the cell values of the "Year" column, searching for four digits within these cell values.(4). The ``value" parameter represents a single cell in the Year column.
\vspace{-7pt}
\begin{equation}
\label{eq:regex_example}
\begin{split}
&\text{jython: import re} \\ 
&\text{match = re.search(r`\textbackslash b\textbackslash d\{4\}\textbackslash b', value)} \\ 
&\text{if match:} \\ 
&\quad \text{return match.group(0)}
\end{split}
\end{equation}

\subsection{Predicted Workflows Analysis}

\begin{figure*}[!bth] 
\centering 
\includegraphics[width=\textwidth]{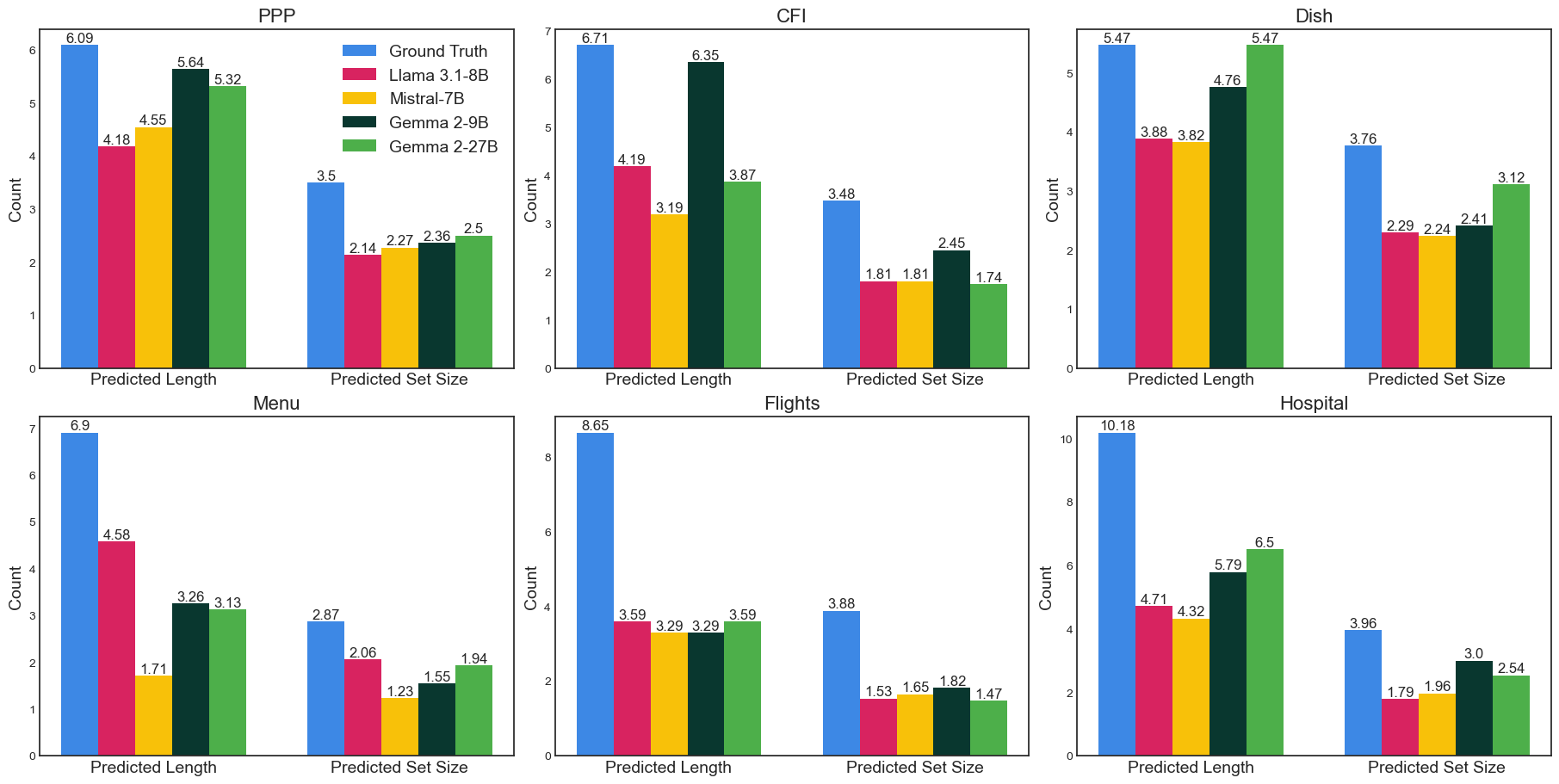} 
\caption{\revise{Workflow Length Across Six Datasets: Operation List Length refers to the total count of steps in each workflow, while Operation Set Length denotes the number of distinct operations within the workflow.}}
\label{fig:op_length}
\end{figure*}

Figure \ref{fig:op_length} illustrates different workflow styles across models. Compared to ground truth workflows, predicted workflows are consistently shorter across all datasets. Similarly, LLM-generated operation sets are smaller than the ground truth. The shorter workflows reflect LLMs' ability to generate more concise and efficient operations for the same data quality issues. A smaller operation set size showcases the models' varying ability to learn and prioritize different operations, reflecting differences in how they optimize workflow generation. 

\vspace{-10pt}
\subsection{Experiment setting}\label{appendix:exp_hyperparams}

The model parameters for predicting the next operation are as follows: \emph{temperature} is set to 0.1, \emph{num\_predict} is -1, \emph{top\_k} is 60, \emph{top\_p} is 0.95, and \emph{mirostat} is set to 1. If the LLMs fail to choose the operation, the \emph{temperature} will be adjusted to 0.3, while the other parameters remain unchanged. The parameters for arguments generation we use in \textit{mass\_edit} are: \emph{temperature} is set to 0.2, with the same stop value.

\vspace{-8pt}
\subsection{Overall Performance Across Datasets} \label{appendix:exp_results}

\begin{table*}[!ht]
\centering
\caption{Overall Performance Results Across Datasets: \{PPP, CFI, Dish, Menu, Hospital, Flights\}. The highest results are \textbf{bolded}.}
\vspace{-0.3cm}
\label{tab:dataset_perf}
\resizebox{\textwidth}{!}{%
\begin{tabular}{cccccccccc}
\hline
\multicolumn{10}{c}{\textbf{PPP}} \\ \hline
\multicolumn{1}{c|}{\multirow{2}{*}{\textbf{Model}}} &
  \multicolumn{4}{c|}{\textbf{Answer Dimension}} &
  \multicolumn{1}{c|}{\textbf{Column Dimension}} &
  \multicolumn{4}{c}{\textbf{Workflow Dimension}} \\ \cline{2-10} 
\multicolumn{1}{c|}{} &
  \multicolumn{1}{c|}{\textbf{Precision}} &
  \multicolumn{1}{c|}{\textbf{Recall}} &
  \multicolumn{1}{c|}{\textbf{F1}} &
  \multicolumn{1}{c|}{\textbf{Similarity}} &
  \multicolumn{1}{c|}{\textbf{Ratio}} &
  \multicolumn{1}{c|}{\textbf{Exact Match}} &
  \multicolumn{1}{c|}{\textbf{Precision}} &
  \multicolumn{1}{c|}{\textbf{Recall}} &
  \textbf{F1} \\ \hline
\multicolumn{1}{c|}{Baseline (Raw Tables)} &
  \multicolumn{1}{c|}{0.3318} &
  \multicolumn{1}{c|}{0.3332} &
  \multicolumn{1}{c|}{0.3309} &
  \multicolumn{1}{c|}{0.5335} &
  \multicolumn{1}{c|}{0.5962} &
  \multicolumn{1}{c|}{-} &
  \multicolumn{1}{c|}{-} &
  \multicolumn{1}{c|}{-} &
  - \\ \hline
\multicolumn{1}{c|}{Llama 3.1-8b (DP)} &
  \multicolumn{1}{c|}{0.3334} &
  \multicolumn{1}{c|}{0.3359} &
  \multicolumn{1}{c|}{0.3329} &
  \multicolumn{1}{c|}{0.5414} &
  \multicolumn{1}{c|}{0.6272} &
  \multicolumn{1}{c|}{0.2273} &
  \multicolumn{1}{c|}{0.6364} &
  \multicolumn{1}{c|}{0.3735} &
  0.4364\\ \hline
\multicolumn{1}{c|}{Llama 3.1-8b (AutoDCWorkflow)} &
  \multicolumn{1}{c|}{0.5812} &
  \multicolumn{1}{c|}{0.5332} &
  \multicolumn{1}{c|}{0.5449} &
  \multicolumn{1}{c|}{0.7138} &
  \multicolumn{1}{c|}{0.8178} &
  \multicolumn{1}{c|}{0.2728} &
  \multicolumn{1}{c|}{\textbf{0.9583}} &
  \multicolumn{1}{c|}{0.6652} &
  0.7559 \\ \hline
\multicolumn{1}{c|}{Mistral-7b (DP)} &
  \multicolumn{1}{c|}{0.3334} &
  \multicolumn{1}{c|}{0.3359} &
  \multicolumn{1}{c|}{0.3329} &
  \multicolumn{1}{c|}{0.5401} &
  \multicolumn{1}{c|}{0.6178} &
  \multicolumn{1}{c|}{0} &
  \multicolumn{1}{c|}{0.3182} &
  \multicolumn{1}{c|}{0.1136} &
  0.1626\\ \hline
\multicolumn{1}{c|}{Mistral-7b (AutoDCWorkflow)} &
  \multicolumn{1}{c|}{0.3182} &
  \multicolumn{1}{c|}{0.3225} &
  \multicolumn{1}{c|}{0.3191} &
  \multicolumn{1}{c|}{0.5387} &
  \multicolumn{1}{c|}{0.6364} &
  \multicolumn{1}{c|}{0.1818} &
  \multicolumn{1}{c|}{0.8470} &
  \multicolumn{1}{c|}{0.5864} &
  0.6604 \\ \hline
\multicolumn{1}{c|}{Gemma2-9b (DP)} &
  \multicolumn{1}{c|}{0.3225} &
  \multicolumn{1}{c|}{0.3287} &
  \multicolumn{1}{c|}{0.3242} &
  \multicolumn{1}{c|}{0.5420} &
  \multicolumn{1}{c|}{0.6375} &
  \multicolumn{1}{c|}{0.0909} &
  \multicolumn{1}{c|}{0.3864} &
  \multicolumn{1}{c|}{0.1826} &
  0.2281\\ \hline
\multicolumn{1}{c|}{Gemma2-9b (AutoDCWorkflow)} &
  \multicolumn{1}{c|}{0.4060} &
  \multicolumn{1}{c|}{0.4133} &
  \multicolumn{1}{c|}{0.4079} &
  \multicolumn{1}{c|}{0.6247} &
  \multicolumn{1}{c|}{0.7098} &
  \multicolumn{1}{c|}{\textbf{0.3182}} &
  \multicolumn{1}{c|}{0.9432} &
  \multicolumn{1}{c|}{\textbf{0.6818}} &
 \textbf{0.7661} \\ \hline
\multicolumn{1}{c|}{Gemma2-27b (DP)} &
  \multicolumn{1}{c|}{0.3225} &
  \multicolumn{1}{c|}{0.3287} &
  \multicolumn{1}{c|}{0.3242} &
  \multicolumn{1}{c|}{0.5420} &
  \multicolumn{1}{c|}{0.6375} &
  \multicolumn{1}{c|}{0.0455} &
  \multicolumn{1}{c|}{0.7955} &
  \multicolumn{1}{c|}{0.3197} &
  0.4318\\ \hline
\multicolumn{1}{c|}{Gemma2-27b (AutoDCWorkflow)} &
  \multicolumn{1}{c|}{\textbf{0.7373}} &
  \multicolumn{1}{c|}{\textbf{0.7259}} &
  \multicolumn{1}{c|}{\textbf{0.7273}} &
  \multicolumn{1}{c|}{\textbf{0.8647}}&
  \multicolumn{1}{c|}{\textbf{0.9038}} &
  \multicolumn{1}{c|}{0.1818} &
  \multicolumn{1}{c|}{0.8598} &
  \multicolumn{1}{c|}{0.6735} &
  0.7285 \\ \hline\hline
\multicolumn{10}{c}{\textbf{CFI}} \\ \hline
\multicolumn{1}{c|}{\multirow{2}{*}{\textbf{Model}}} &
  \multicolumn{4}{c|}{\textbf{Answer Dimension}} &
  \multicolumn{1}{c|}{\textbf{Column Dimension}} &
  \multicolumn{4}{c}{\textbf{Workflow Dimension}} \\ \cline{2-10} 
\multicolumn{1}{c|}{} &
  \multicolumn{1}{c|}{\textbf{Precision}} &
  \multicolumn{1}{c|}{\textbf{Recall}} &
  \multicolumn{1}{c|}{\textbf{F1}} &
  \multicolumn{1}{c|}{\textbf{Similarity}} &
  \multicolumn{1}{c|}{\textbf{Ratio}} &
  \multicolumn{1}{c|}{\textbf{Exact Match}} &
  \multicolumn{1}{c|}{\textbf{Precision}} &
  \multicolumn{1}{c|}{\textbf{Recall}} &
  \textbf{F1} \\ \hline
\multicolumn{1}{c|}{Baseline (Raw Tables)} &
  \multicolumn{1}{c|}{0.2511} &
  \multicolumn{1}{c|}{0.2554} &
  \multicolumn{1}{c|}{0.2318} &
  \multicolumn{1}{c|}{0.4967} &
  \multicolumn{1}{c|}{0.3829} &
  \multicolumn{1}{c|}{-} &
  \multicolumn{1}{c|}{-} &
  \multicolumn{1}{c|}{-} &
  - \\ \hline
\multicolumn{1}{c|}{Llama 3.1-8b (DP)} &
  \multicolumn{1}{c|}{0.2357} &
  \multicolumn{1}{c|}{0.2554} &
  \multicolumn{1}{c|}{0.2248} &
  \multicolumn{1}{c|}{0.4610} &
  \multicolumn{1}{c|}{0.5295} &
  \multicolumn{1}{c|}{0.0323} &
  \multicolumn{1}{c|}{0.1935} &
  \multicolumn{1}{c|}{0.0952} &
  0.1184\\ \hline
\multicolumn{1}{c|}{Llama 3.1-8b (AutoDCWorkflow)} &
  \multicolumn{1}{c|}{0.6444} &
  \multicolumn{1}{c|}{0.6680} &
  \multicolumn{1}{c|}{0.6484} &
  \multicolumn{1}{c|}{0.7705} &
  \multicolumn{1}{c|}{0.7238} &
  \multicolumn{1}{c|}{0.0645} &
  \multicolumn{1}{c|}{\textbf{0.9409}} &
  \multicolumn{1}{c|}{0.5124} &
  \textbf{0.6382} \\ \hline
\multicolumn{1}{c|}{Mistral-7b (DP)} &
  \multicolumn{1}{c|}{0.2228} &
  \multicolumn{1}{c|}{0.2366} &
  \multicolumn{1}{c|}{0.2118} &
  \multicolumn{1}{c|}{0.4875} &
  \multicolumn{1}{c|}{0.4157} &
  \multicolumn{1}{c|}{0} &
  \multicolumn{1}{c|}{0.1129} &
  \multicolumn{1}{c|}{0.0290} &
  0.0436\\ \hline
\multicolumn{1}{c|}{Mistral-7b (AutoDCWorkflow)} &
  \multicolumn{1}{c|}{0.3884} &
  \multicolumn{1}{c|}{0.3441} &
  \multicolumn{1}{c|}{0.3496} &
  \multicolumn{1}{c|}{0.5720} &
  \multicolumn{1}{c|}{0.5864} &
  \multicolumn{1}{c|}{0.0323} &
  \multicolumn{1}{c|}{0.7957} &
  \multicolumn{1}{c|}{0.4161} &
  0.5095 \\ \hline
\multicolumn{1}{c|}{Gemma2-9b (DP)} &
  \multicolumn{1}{c|}{0.2357} &
  \multicolumn{1}{c|}{0.2554} &
  \multicolumn{1}{c|}{0.2248} &
  \multicolumn{1}{c|}{0.4610} &
  \multicolumn{1}{c|}{0.5420} &
  \multicolumn{1}{c|}{0} &
  \multicolumn{1}{c|}{0.1935} &
  \multicolumn{1}{c|}{0.0452} &
  0.0727\\ \hline
\multicolumn{1}{c|}{Gemma2-9b (AutoDCWorkflow)} &
  \multicolumn{1}{c|}{0.3568} &
  \multicolumn{1}{c|}{0.3387} &
  \multicolumn{1}{c|}{0.3272} &
  \multicolumn{1}{c|}{0.5289} &
  \multicolumn{1}{c|}{0.5743} &
  \multicolumn{1}{c|}{0.0645} &
  \multicolumn{1}{c|}{0.8081} &
  \multicolumn{1}{c|}{\textbf{0.5452}} &
 0.6178 \\ \hline
\multicolumn{1}{c|}{Gemma2-27b (DP)} &
  \multicolumn{1}{c|}{0.2679} &
  \multicolumn{1}{c|}{0.2715} &
  \multicolumn{1}{c|}{0.2463} &
  \multicolumn{1}{c|}{0.4613} &
  \multicolumn{1}{c|}{0.5528} &
  \multicolumn{1}{c|}{0} &
  \multicolumn{1}{c|}{0.3387} &
  \multicolumn{1}{c|}{0.0935} &
  0.1436\\ \hline
\multicolumn{1}{c|}{Gemma2-27b (AutoDCWorkflow)} &
  \multicolumn{1}{c|}{\textbf{0.6911}} &
  \multicolumn{1}{c|}{\textbf{0.7366}} &
  \multicolumn{1}{c|}{\textbf{0.7026}} &
  \multicolumn{1}{c|}{\textbf{0.7992}} &
  \multicolumn{1}{c|}{\textbf{0.7455}} &
  \multicolumn{1}{c|}{\textbf{0.0968}} &
  \multicolumn{1}{c|}{0.9355} &
  \multicolumn{1}{c|}{0.4844} &
  0.6120 \\ \hline\hline
\multicolumn{10}{c}{\textbf{Dish}} \\ \hline
\multicolumn{1}{c|}{\multirow{2}{*}{\textbf{Model}}} &
  \multicolumn{4}{c|}{\textbf{Answer Dimension}} &
  \multicolumn{1}{c|}{\textbf{Column Dimension}} &
  \multicolumn{4}{c}{\textbf{Workflow Dimension}} \\ \cline{2-10} 
\multicolumn{1}{c|}{} &
  \multicolumn{1}{c|}{\textbf{Precision}} &
  \multicolumn{1}{c|}{\textbf{Recall}} &
  \multicolumn{1}{c|}{\textbf{F1}} &
  \multicolumn{1}{c|}{\textbf{Similarity}} &
  \multicolumn{1}{c|}{\textbf{Ratio}} &
  \multicolumn{1}{c|}{\textbf{Exact Match}} &
  \multicolumn{1}{c|}{\textbf{Precision}} &
  \multicolumn{1}{c|}{\textbf{Recall}} &
  \textbf{F1} \\ \hline
\multicolumn{1}{c|}{Baseline (Raw Tables)} &
  \multicolumn{1}{c|}{0.1471} &
  \multicolumn{1}{c|}{0.1471} &
  \multicolumn{1}{c|}{0.1471} &
  \multicolumn{1}{c|}{0.6771} &
  \multicolumn{1}{c|}{0.4863} &
  \multicolumn{1}{c|}{-} &
  \multicolumn{1}{c|}{-} &
  \multicolumn{1}{c|}{-} &
  - \\ \hline
\multicolumn{1}{c|}{Llama 3.1-8b (DP)} &
  \multicolumn{1}{c|}{0.1471} &
  \multicolumn{1}{c|}{0.1471} &
  \multicolumn{1}{c|}{0.1471} &
  \multicolumn{1}{c|}{0.6399} &
  \multicolumn{1}{c|}{0.5505} &
  \multicolumn{1}{c|}{0.0588} &
  \multicolumn{1}{c|}{0.9118} &
  \multicolumn{1}{c|}{0.2990} &
  0.4294\\ \hline
\multicolumn{1}{c|}{Llama 3.1-8b (AutoDCWorkflow)} &
  \multicolumn{1}{c|}{0.3269} &
  \multicolumn{1}{c|}{0.2992} &
  \multicolumn{1}{c|}{0.3078} &
  \multicolumn{1}{c|}{0.6779} &
  \multicolumn{1}{c|}{0.6684} &
  \multicolumn{1}{c|}{\textbf{0.1176}} &
  \multicolumn{1}{c|}{0.8676} &
  \multicolumn{1}{c|}{0.5147} &
  \textbf{0.6144} \\ \hline
\multicolumn{1}{c|}{Mistral-7b (DP)} &
  \multicolumn{1}{c|}{0.2059} &
  \multicolumn{1}{c|}{0.1711} &
  \multicolumn{1}{c|}{0.1812} &
  \multicolumn{1}{c|}{0.7036} &
  \multicolumn{1}{c|}{0.5224} &
  \multicolumn{1}{c|}{0} &
  \multicolumn{1}{c|}{0.0588} &
  \multicolumn{1}{c|}{0.0147} &
  0.0235\\ \hline
\multicolumn{1}{c|}{Mistral-7b (AutoDCWorkflow)} &
  \multicolumn{1}{c|}{0.2059} &
  \multicolumn{1}{c|}{0.1711} &
  \multicolumn{1}{c|}{0.1812} &
  \multicolumn{1}{c|}{0.5482} &
  \multicolumn{1}{c|}{0.5413} &
  \multicolumn{1}{c|}{0.0588} &
  \multicolumn{1}{c|}{0.7598} &
  \multicolumn{1}{c|}{0.4314} &
  0.5172 \\ \hline
\multicolumn{1}{c|}{Gemma2-9b (DP)} &
  \multicolumn{1}{c|}{0.2059} &
  \multicolumn{1}{c|}{0.1711} &
  \multicolumn{1}{c|}{0.1812} &
  \multicolumn{1}{c|}{0.6664} &
  \multicolumn{1}{c|}{0.5899} &
  \multicolumn{1}{c|}{0.0} &
  \multicolumn{1}{c|}{0.7941} &
  \multicolumn{1}{c|}{0.2402} &
  0.3608\\ \hline
\multicolumn{1}{c|}{Gemma2-9b (AutoDCWorkflow)} &
  \multicolumn{1}{c|}{0.3235} &
  \multicolumn{1}{c|}{0.2888} &
  \multicolumn{1}{c|}{0.2989} &
  \multicolumn{1}{c|}{0.6884} &
  \multicolumn{1}{c|}{0.6066} &
  \multicolumn{1}{c|}{0.0588} &
  \multicolumn{1}{c|}{0.8088}&
  \multicolumn{1}{c|}{0.5049} &
  0.5889\\ \hline
\multicolumn{1}{c|}{Gemma2-27b (DP)} &
  \multicolumn{1}{c|}{0.2059} &
  \multicolumn{1}{c|}{0.1711} &
  \multicolumn{1}{c|}{0.1812} &
  \multicolumn{1}{c|}{0.6664} &
  \multicolumn{1}{c|}{0.5899} &
  \multicolumn{1}{c|}{0.0588} &
  \multicolumn{1}{c|}{\textbf{1.0000}} &
  \multicolumn{1}{c|}{0.3725} &
  0.5196\\ \hline
\multicolumn{1}{c|}{Gemma2-27b (AutoDCWorkflow)} &
  \multicolumn{1}{c|}{\textbf{0.6647}} &
  \multicolumn{1}{c|}{\textbf{0.5977}} &
  \multicolumn{1}{c|}{\textbf{0.6178}} &
  \multicolumn{1}{c|}{\textbf{0.7774}} &
  \multicolumn{1}{c|}{\textbf{0.7922}} &
  \multicolumn{1}{c|}{0} &
  \multicolumn{1}{c|}{0.6745} &
  \multicolumn{1}{c|}{\textbf{0.5833}} &
  0.5999 \\ \hline\hline
\multicolumn{10}{c}{\textbf{Menu}} \\ \hline
\multicolumn{1}{c|}{\multirow{2}{*}{\textbf{Model}}} &
  \multicolumn{4}{c|}{\textbf{Answer Dimension}} &
  \multicolumn{1}{c|}{\textbf{Column Dimension}} &
  \multicolumn{4}{c}{\textbf{Workflow Dimension}} \\ \cline{2-10} 
\multicolumn{1}{c|}{} &
  \multicolumn{1}{c|}{\textbf{Precision}} &
  \multicolumn{1}{c|}{\textbf{Recall}} &
  \multicolumn{1}{c|}{\textbf{F1}} &
  \multicolumn{1}{c|}{\textbf{Similarity}} &
  \multicolumn{1}{c|}{\textbf{Ratio}} &
  \multicolumn{1}{c|}{\textbf{Exact Match}} &
  \multicolumn{1}{c|}{\textbf{Precision}} &
  \multicolumn{1}{c|}{\textbf{Recall}} &
  \textbf{F1} \\ \hline
\multicolumn{1}{c|}{Baseline (Raw Tables)} &
  \multicolumn{1}{c|}{0.4172} &
  \multicolumn{1}{c|}{0.4353} &
  \multicolumn{1}{c|}{0.3851} &
  \multicolumn{1}{c|}{0.4973} &
  \multicolumn{1}{c|}{0.5104} &
  \multicolumn{1}{c|}{-} &
  \multicolumn{1}{c|}{-} &
  \multicolumn{1}{c|}{-} &
  - \\ \hline
\multicolumn{1}{c|}{Llama 3.1-8b (DP)} &
  \multicolumn{1}{c|}{0.4672} &
  \multicolumn{1}{c|}{0.4676} &
  \multicolumn{1}{c|}{0.4282} &
  \multicolumn{1}{c|}{0.5464} &
  \multicolumn{1}{c|}{0.6124} &
  \multicolumn{1}{c|}{0.0323} &
  \multicolumn{1}{c|}{0.1935} &
  \multicolumn{1}{c|}{0.1048} &
  0.1301\\ \hline
\multicolumn{1}{c|}{Llama 3.1-8b (AutoDCWorkflow)} &
  \multicolumn{1}{c|}{0.4688} &
  \multicolumn{1}{c|}{0.4646} &
  \multicolumn{1}{c|}{0.4408} &
  \multicolumn{1}{c|}{0.5622} &
  \multicolumn{1}{c|}{0.6542} &
  \multicolumn{1}{c|}{\textbf{0.1935}} &
  \multicolumn{1}{c|}{\textbf{0.8710}} &
  \multicolumn{1}{c|}{\textbf{0.6231}} &
  \textbf{0.7011} \\ \hline
\multicolumn{1}{c|}{Mistral-7b (DP)} &
  \multicolumn{1}{c|}{0.4591} &
  \multicolumn{1}{c|}{0.4622} &
  \multicolumn{1}{c|}{0.4136} &
  \multicolumn{1}{c|}{0.5165} &
  \multicolumn{1}{c|}{0.5455} &
  \multicolumn{1}{c|}{0} &
  \multicolumn{1}{c|}{0} &
  \multicolumn{1}{c|}{0} &
  0\\ \hline
\multicolumn{1}{c|}{Mistral-7b (AutoDCWorkflow)} &
  \multicolumn{1}{c|}{0.5000} &
  \multicolumn{1}{c|}{0.4943} &
  \multicolumn{1}{c|}{0.4632} &
  \multicolumn{1}{c|}{0.5438} &
  \multicolumn{1}{c|}{0.5984} &
  \multicolumn{1}{c|}{0.0645} &
  \multicolumn{1}{c|}{0.4892} &
  \multicolumn{1}{c|}{0.2527} &
  0.3154 \\ \hline
\multicolumn{1}{c|}{Gemma2-9b (DP)} &
  \multicolumn{1}{c|}{0.4661} &
  \multicolumn{1}{c|}{0.4717} &
  \multicolumn{1}{c|}{0.4325} &
  \multicolumn{1}{c|}{0.5246} &
  \multicolumn{1}{c|}{0.5675} &
  \multicolumn{1}{c|}{0} &
  \multicolumn{1}{c|}{0} &
  \multicolumn{1}{c|}{0} &
  0\\ \hline
\multicolumn{1}{c|}{Gemma2-9b (AutoDCWorkflow)} &
  \multicolumn{1}{c|}{0.4685} &
  \multicolumn{1}{c|}{0.4717} &
  \multicolumn{1}{c|}{0.4345} &
  \multicolumn{1}{c|}{0.5497} &
  \multicolumn{1}{c|}{0.6666} &
  \multicolumn{1}{c|}{0.0645} &
  \multicolumn{1}{c|}{0.8226} &
  \multicolumn{1}{c|}{0.4645} &
  0.5661\\ \hline
\multicolumn{1}{c|}{Gemma2-27b (DP)} &
  \multicolumn{1}{c|}{0.4661} &
  \multicolumn{1}{c|}{0.4717} &
  \multicolumn{1}{c|}{0.4325} &
  \multicolumn{1}{c|}{0.5265} &
  \multicolumn{1}{c|}{0.5675} &
  \multicolumn{1}{c|}{0} &
  \multicolumn{1}{c|}{0} &
  \multicolumn{1}{c|}{0} &
  0\\ \hline
\multicolumn{1}{c|}{Gemma2-27b (AutoDCWorkflow)} &
  \multicolumn{1}{c|}{\textbf{0.6620}} &
  \multicolumn{1}{c|}{\textbf{0.6213}} &
  \multicolumn{1}{c|}{\textbf{0.6167}} &
  \multicolumn{1}{c|}{\textbf{0.6849}} &
  \multicolumn{1}{c|}{\textbf{0.7042}} &
  \multicolumn{1}{c|}{0.1613} &
  \multicolumn{1}{c|}{0.8172} &
  \multicolumn{1}{c|}{0.5866} &
  0.657 \\ \hline\hline
\multicolumn{10}{c}{\textbf{Hospital}} \\ \hline
\multicolumn{1}{c|}{\multirow{2}{*}{\textbf{Model}}} &
  \multicolumn{4}{c|}{\textbf{Answer Dimension}} &
  \multicolumn{1}{c|}{\textbf{Column Dimension}} &
  \multicolumn{4}{c}{\textbf{Workflow Dimension}} \\ \cline{2-10} 
\multicolumn{1}{c|}{} &
  \multicolumn{1}{c|}{\textbf{Precision}} &
  \multicolumn{1}{c|}{\textbf{Recall}} &
  \multicolumn{1}{c|}{\textbf{F1}} &
  \multicolumn{1}{c|}{\textbf{Similarity}} &
  \multicolumn{1}{c|}{\textbf{Ratio}} &
  \multicolumn{1}{c|}{\textbf{Exact Match}} &
  \multicolumn{1}{c|}{\textbf{Precision}} &
  \multicolumn{1}{c|}{\textbf{Recall}} &
  \textbf{F1} \\ \hline
\multicolumn{1}{c|}{Baseline (Raw Tables)} &
  \multicolumn{1}{c|}{0.1045} &
  \multicolumn{1}{c|}{0.1142} &
  \multicolumn{1}{c|}{0.1056} &
  \multicolumn{1}{c|}{0.3498} &
  \multicolumn{1}{c|}{0.4533} &
  \multicolumn{1}{c|}{-} &
  \multicolumn{1}{c|}{-} &
  \multicolumn{1}{c|}{-} &
  - \\ \hline
\multicolumn{1}{c|}{Llama 3.1-8b (DP)} &
  \multicolumn{1}{c|}{0.2098} &
  \multicolumn{1}{c|}{0.2120} &
  \multicolumn{1}{c|}{0.2070} &
  \multicolumn{1}{c|}{0.4466} &
  \multicolumn{1}{c|}{0.6979} &
  \multicolumn{1}{c|}{0} &
  \multicolumn{1}{c|}{0.1786} &
  \multicolumn{1}{c|}{0.0518} &
  0.0793\\ \hline
\multicolumn{1}{c|}{Llama 3.1-8b (AutoDCWorkflow)} &
  \multicolumn{1}{c|}{\textbf{0.5759}} &
  \multicolumn{1}{c|}{0.5532} &
  \multicolumn{1}{c|}{0.5622} &
  \multicolumn{1}{c|}{0.7849} &
  \multicolumn{1}{c|}{0.8911} &
  \multicolumn{1}{c|}{0} &
  \multicolumn{1}{c|}{\textbf{1.0000}} &
  \multicolumn{1}{c|}{0.4583} &
  0.6187 \\ \hline
\multicolumn{1}{c|}{Mistral-7b (DP)} &
  \multicolumn{1}{c|}{0.1045} &
  \multicolumn{1}{c|}{0.1142} &
  \multicolumn{1}{c|}{0.1056} &
  \multicolumn{1}{c|}{0.3498} &
  \multicolumn{1}{c|}{0.4533} &
  \multicolumn{1}{c|}{0} &
  \multicolumn{1}{c|}{0} &
  \multicolumn{1}{c|}{0} &
  0\\ \hline
\multicolumn{1}{c|}{Mistral-7b (AutoDCWorkflow)} &
  \multicolumn{1}{c|}{0.2634} &
  \multicolumn{1}{c|}{0.2656} &
  \multicolumn{1}{c|}{0.2606} &
  \multicolumn{1}{c|}{0.4750} &
  \multicolumn{1}{c|}{0.6711} &
  \multicolumn{1}{c|}{0} &
  \multicolumn{1}{c|}{0.9226} &
  \multicolumn{1}{c|}{0.4536} &
  0.5914 \\ \hline
\multicolumn{1}{c|}{Gemma2-9b (DP)} &
  \multicolumn{1}{c|}{0.2098} &
  \multicolumn{1}{c|}{0.2120} &
  \multicolumn{1}{c|}{0.2070} &
  \multicolumn{1}{c|}{0.4466} &
  \multicolumn{1}{c|}{0.6890} &
  \multicolumn{1}{c|}{0} &
  \multicolumn{1}{c|}{0.1071} &
  \multicolumn{1}{c|}{0.0387} &
  0.0560\\ \hline
\multicolumn{1}{c|}{Gemma2-9b (AutoDCWorkflow)} &
  \multicolumn{1}{c|}{0.2446} &
  \multicolumn{1}{c|}{0.2375} &
  \multicolumn{1}{c|}{0.2385} &
  \multicolumn{1}{c|}{0.5084} &
  \multicolumn{1}{c|}{0.7762} &
  \multicolumn{1}{c|}{\textbf{0.1429}} &
  \multicolumn{1}{c|}{0.9881} &
  \multicolumn{1}{c|}{\textbf{0.7440}} &
  \textbf{0.8392}\\ \hline
\multicolumn{1}{c|}{Gemma2-27b (DP)} &
  \multicolumn{1}{c|}{0.2098} &
  \multicolumn{1}{c|}{0.2120} &
  \multicolumn{1}{c|}{0.2070} &
  \multicolumn{1}{c|}{0.4466} &
  \multicolumn{1}{c|}{0.6979} &
  \multicolumn{1}{c|}{0} &
  \multicolumn{1}{c|}{0.0595} &
  \multicolumn{1}{c|}{0.0351} &
  0.0440\\ \hline
\multicolumn{1}{c|}{Gemma2-27b (AutoDCWorkflow)} &
  \multicolumn{1}{c|}{0.5750} &
  \multicolumn{1}{c|}{\textbf{0.5723}} &
  \multicolumn{1}{c|}{\textbf{0.5736}} &
  \multicolumn{1}{c|}{\textbf{0.8258}} &
  \multicolumn{1}{c|}{\textbf{0.9393}} &
  \multicolumn{1}{c|}{0.0357} &
  \multicolumn{1}{c|}{0.9524} &
  \multicolumn{1}{c|}{0.6113} &
  0.7323 \\ \hline\hline
\multicolumn{10}{c}{\textbf{Flights}} \\ \hline
\multicolumn{1}{c|}{\multirow{2}{*}{\textbf{Model}}} &
  \multicolumn{4}{c|}{\textbf{Answer Dimension}} &
  \multicolumn{1}{c|}{\textbf{Column Dimension}} &
  \multicolumn{4}{c}{\textbf{Workflow Dimension}} \\ \cline{2-10} 
\multicolumn{1}{c|}{} &
  \multicolumn{1}{c|}{\textbf{Precision}} &
  \multicolumn{1}{c|}{\textbf{Recall}} &
  \multicolumn{1}{c|}{\textbf{F1}} &
  \multicolumn{1}{c|}{\textbf{Similarity}} &
  \multicolumn{1}{c|}{\textbf{Ratio}} &
  \multicolumn{1}{c|}{\textbf{Exact Match}} &
  \multicolumn{1}{c|}{\textbf{Precision}} &
  \multicolumn{1}{c|}{\textbf{Recall}} &
  \textbf{F1} \\ \hline
\multicolumn{1}{c|}{Baseline (Raw Tables)} &
  \multicolumn{1}{c|}{0.0895} &
  \multicolumn{1}{c|}{0.0557} &
  \multicolumn{1}{c|}{0.0625} &
  \multicolumn{1}{c|}{0.2668} &
  \multicolumn{1}{c|}{0.0586} &
  \multicolumn{1}{c|}{-} &
  \multicolumn{1}{c|}{-} &
  \multicolumn{1}{c|}{-} &
  - \\ \hline
\multicolumn{1}{c|}{Llama 3.1-8b (DP)} &
  \multicolumn{1}{c|}{0.4567} &
  \multicolumn{1}{c|}{0.2823} &
  \multicolumn{1}{c|}{0.3365} &
  \multicolumn{1}{c|}{0.4111} &
  \multicolumn{1}{c|}{0.4238} &
  \multicolumn{1}{c|}{0} &
  \multicolumn{1}{c|}{0.5000} &
  \multicolumn{1}{c|}{0.1392} &
  0.2157\\ \hline
\multicolumn{1}{c|}{Llama 3.1-8b (AutoDCWorkflow)} &
  \multicolumn{1}{c|}{0.5487} &
  \multicolumn{1}{c|}
  {\textbf{0.4517}} &
  \multicolumn{1}{c|}{\textbf{0.4879}} &
  \multicolumn{1}{c|}{0.5169} &
  \multicolumn{1}{c|}{\textbf{0.7523}} &
  \multicolumn{1}{c|}{0} &
  \multicolumn{1}{c|}{\textbf{0.9706}} &
  \multicolumn{1}{c|}{0.3833} &
  0.5349 \\ \hline
\multicolumn{1}{c|}{Mistral-7b (DP)} &
  \multicolumn{1}{c|}{0.4947} &
  \multicolumn{1}{c|}{0.3387} &
  \multicolumn{1}{c|}{0.3876} &
  \multicolumn{1}{c|}{0.4525} &
  \multicolumn{1}{c|}{0.4269} &
  \multicolumn{1}{c|}{0} &
  \multicolumn{1}{c|}{0.3529} &
  \multicolumn{1}{c|}{0.1029} &
  0.1574\\ \hline
\multicolumn{1}{c|}{Mistral-7b (AutoDCWorkflow)} &
  \multicolumn{1}{c|}{0.5812} &
  \multicolumn{1}{c|}{0.4000} &
  \multicolumn{1}{c|}{0.4593} &
  \multicolumn{1}{c|}{0.4885} &
  \multicolumn{1}{c|}{0.5794} &
  \multicolumn{1}{c|}{0} &
  \multicolumn{1}{c|}{0.8529} &
  \multicolumn{1}{c|}{0.3578} &
  0.4927 \\ \hline
\multicolumn{1}{c|}{Gemma2-9b (DP)} &
  \multicolumn{1}{c|}{\textbf{0.6635}} &
  \multicolumn{1}{c|}{0.3894} &
  \multicolumn{1}{c|}{0.4618} &
  \multicolumn{1}{c|}{0.4779} &
  \multicolumn{1}{c|}{0.6063} &
  \multicolumn{1}{c|}{0} &
  \multicolumn{1}{c|}{0.6471} &
  \multicolumn{1}{c|}{0.1745} &
  0.2728\\ \hline
\multicolumn{1}{c|}{Gemma2-9b (AutoDCWorkflow)} &
  \multicolumn{1}{c|}{0.6047} &
  \multicolumn{1}{c|}{0.3502} &
  \multicolumn{1}{c|}{0.4253} &
  \multicolumn{1}{c|}{0.4660} &
  \multicolumn{1}{c|}{0.6208} &
  \multicolumn{1}{c|}{0} &
  \multicolumn{1}{c|}{0.9412} &
  \multicolumn{1}{c|}{\textbf{0.4186}} &
  \textbf{0.5542}\\ \hline
\multicolumn{1}{c|}{Gemma2-27b (DP)} &
  \multicolumn{1}{c|}{\textbf{0.6635}} &
  \multicolumn{1}{c|}{0.4090} &
  \multicolumn{1}{c|}{0.4842} &
  \multicolumn{1}{c|}{0.5081} &
  \multicolumn{1}{c|}{0.6260} &
  \multicolumn{1}{c|}{0} &
  \multicolumn{1}{c|}{0.7941} &
  \multicolumn{1}{c|}{0.2206} &
  0.3395\\ \hline
\multicolumn{1}{c|}{Gemma2-27b (AutoDCWorkflow)} &
  \multicolumn{1}{c|}{\textbf{0.6635}} &
  \multicolumn{1}{c|}{0.4090} &
  \multicolumn{1}{c|}{0.4842} &
  \multicolumn{1}{c|}{\textbf{0.6586}} &
  \multicolumn{1}{c|}{0.6194} &
  \multicolumn{1}{c|}{0} &
  \multicolumn{1}{c|}{0.9412} &
  \multicolumn{1}{c|}{0.3431} &
  0.4908 \\ \hline
\end{tabular}%
}
\vspace{-0.5cm}
\end{table*}
\revise{We summarize the detailed performance of each model across the six datasets in Table~\ref{tab:dataset_perf}. The results highlight how well different LLMs perform under varying domain characteristics and data quality challenges.}



\vspace{-10pt}
\subsection{Prompt Templates} 
\label{sec:appendix_eod}
\begin{tcolorbox}[colback=gray!5!white, colframe=gray!50!black]
\small
\captionsetup{type=table}
\caption{Generating Data Quality Report.}
\label{tb:prompt_quality}
You are an expert in data cleaning theory and practices. You can recognize whether the current data (i.e., column and cell values) is clean enough to fulfill the objective. The evaluation process for deciding if the pipeline can end (Flag = True): \\
(1) Profile the column at both schema and instance level: Is the column name meaningful? What is the data distribution? Are values clearly represented? \\
(2) Assess four quality dimensions: \\
\textbf{Accuracy:} Are there obvious errors, inconsistencies, or biases? \\
\textbf{Relevance:} Is the column needed to address the purpose? \\
\textbf{Completeness:} Are there enough instances and minimal missing values? \\
\textbf{Conciseness:} Are spellings standardized? No different forms for same meaning? \\
(3) Return True/False for each dimension.  \\
Only when all are True, return Flag as True; otherwise, return False and continue cleaning. \\
\textbf{Example}
\vspace{-3pt}
{\ttfamily
\[
\begin{array}{| l | l | l | l |}
    \hline
    \textbf{LoanAmount} & \textbf{City} & \textbf{State} & \textbf{Zip} \\
    \hline
    30333 & Honolulu & HI & 96814 \\
    149900 & Honolulu & HI &  \\
    148100 & Honolulu & HI & 96814 \\
    334444 &  & IL &  \\
    120 & Urbana & IL & 61802 \\
    100000 & Chicago & IL &  \\
    1000. & Champaign & IL & 61820 \\
    \hline
\end{array}
\]
}

Purpose: Figure out how many cities are in the table.  \\
Flag: True \\  
Target column: City  \\
Explanations: Accuracy: True (correct spellings for the same city names and same format in column City) Relevance: True (column City is relevant to the Purpose) Completeness: NA (with minor number of missing values in column City but it (1/7) can be ignored) Conciseness: True (incorrect variations do not exist in column City) Since there are no concerns (True or NA) with the quality dimensions, I will return True.\\
\dots

\label{tb:prompt_eod}
\end{tcolorbox}

\begin{tcolorbox}[colback=gray!5!white, colframe=gray!50!black]
\small
\captionsetup{type=table} 
\caption{Prompt introducing data cleaning operations.}
You are an expert in data cleaning and are able to choose appropriate Operations to prepare the table in good format before addressing the Purpose. Note that the operation chosen should aim at making the data be in a better shape that can be used for the purpose instead of addressing the purpose directly.
The Operations pool you can choose from contains \textit{upper}, \textit{trim}, \textit{mass\_edit},  \textit{regexr\_transform}, \textit{numeric}, and \textit{date}.

Available example demos to learn the data cleaning operations are as follows:

1. upper: The upper function is used to convert all cell values in a column that are strings into uppercase, fixing formatting error for strings. This is particularly useful for standardizing data, especially when dealing with categorical variables, to ensure consistency in text representation. Standardizing case can help avoid issues with duplicate entries that differ only in capitalization. For example,\\
\ttfamily 
$
\begin{array}{| l | l | l |}
    \hline
    \textbf{id} & \textbf{neighbourhood} & \textbf{room type}\\
    \hline
    46154 & Ohare & Entire home/ apt\\
    6715 & OHARE & Entire home/ apt \\
    228273 & ohare & Private\ room \\
    \hline
\end{array}
$
\\
Purpose: Return room types that are located near OHARE.\\
Target column: neighbourhood \\
Selected Operation: upper \\
Explanation: Improve conciseness: The format of cell values in column neighbourhood are inconsistent(mixed with different formats). Therefore, We use upper on column "neighbourhood" to make the format consistent as Uppercase.\\
Output: OHARE | OHARE | OHARE \\
\dots
\label{tb:prompt_ops}
\end{tcolorbox}

\begin{tcolorbox}[colback=gray!5!white, colframe=gray!50!black]
\small
\captionsetup{type=table} 
\caption{Prompt Template: Column Selection Based on Purpose}
\label{tb:prompt_col}
\textbf{Instruction:}  
Select relevant column(s) from a table given its metadata and a natural language purpose.\\
\textbf{Example:} 
\begin{lstlisting}[basicstyle=\ttfamily\small, breaklines=true, frame=none]
{
  "table_caption": "List of largest airlines in Central America & the Caribbean",
  "columns": ["rank", "airline", "country", "fleet size", "remarks"],
  "table_column_priority": [
    ["rank", "1", "2", "3"],
    ["airline", "Caribbean Airlines", "LIAT", "Cubana de Aviaci\u00e3 n"],
    ["country", "Trinidad and Tobago", "Antigua and Barbuda", "Cuba"],
    ["fleet size", "22", "17", "14"],
    ["remarks", "Largest airline in the Caribbean", "Second largest airline", "Operational since 1929"]
  ],
}
\end{lstlisting}
\textbf{Purpose}: "How many countries are involved?"\\
\textbf{Selected columns}: ```['country']```\\
\textbf{Explanations}:\\
similar words link to columns :
countries -> country.
\end{tcolorbox}

\end{document}